\documentclass[11pt]{article}

\usepackage[preprint]{acl}

\usepackage{times}
\usepackage{latexsym}

\usepackage[T1]{fontenc}

\usepackage[utf8]{inputenc}

\usepackage{microtype}

\usepackage{inconsolata}

\usepackage{graphicx}
\usepackage{amsmath}
\usepackage{multirow}
\usepackage{placeins}
\newcommand*{\demourl}{https://anonymous-user428.github.io/evo-speaker-mem-demo}

\title{From Voiceprints to User Entities: Online Speaker-Memory Construction for Multi-User Spoken Agents}

\author{
  \textbf{Yucong Zhang\textsuperscript{3,1,2,$\ast$}},
  \textbf{Cancan Li\textsuperscript{3,2,}\thanks{Equal Contribution. Work done during an internship at Meituan.}},
  \textbf{Bo Xiao\textsuperscript{2}},
  \textbf{Ke Zeng\textsuperscript{2}},
  \textbf{Ming Li\textsuperscript{1,}\thanks{Corresponding author: Ming Li}}
\\
  \textsuperscript{1}School of Artificial Intelligence, The Chinese University of Hong Kong, Shenzhen, \\
  \textsuperscript{2}Meituan,
  \textsuperscript{3}School of Computer Science, Wuhan University
\\
  \small{
    \{yucong.zhang, cancan.li\}@whu.edu.cn, mingli369@cuhk.edu.cn
    }
}

\begin{document}
\maketitle
\begin{abstract}
Personalized spoken agents must identify speakers before accessing user-specific memory, yet memory benchmarks typically assume correctly attributed histories. We argue that voiceprints are observations of identity, not identities themselves, and formulate online speaker-memory construction to maintain persistent user entities as acoustic evidence evolves. We introduce EvoSpeakerMem, a training-free layer for causal speaker routing, selective acoustic-state updates, and identity-partitioned memory writes and retrieval through a replaceable text-memory backend. We also construct LoCoMo-ID, a spoken-memory benchmark with hidden speaker identities, speaker-sensitive queries, simulated recording-condition shifts, and four-speaker identity-confusion tests. On longitudinal SMIIP-TV, genuine-probe rejection under a frozen near-enrollment threshold rises from 29.8\% to 55.6\% over approximately one year. On LoCoMo-ID speaker-sensitive queries, gated updating improves identity-gated QA over fixed representations by 3.86 and 11.62 percentage points under clean and shifted conditions, respectively, with no wrong-user authorizations. The advantage persists in the four-speaker test. Yet reducing fragmentation and improving QA can coexist with incorrect cross-user links, revealing a fragmentation--contamination trade-off that QA alone cannot capture.
\end{abstract}

\section{Introduction}

Personalized spoken agents rely on user profiles and interaction histories to remain consistent across turns and sessions~\cite{zhang-etal-2018-personalizing,xu-etal-2022-long}. In shared spoken environments, however, maintaining such memory first requires deciding \emph{whose} history each incoming observation belongs to. Speaker attribution is therefore a prerequisite for personalized memory construction. An identity error is stateful: treating a returning user as new fragments one person's history across profiles, whereas confusing two users contaminates one profile with another user's information and can affect subsequent personalized responses.

Recent audio large language models~(Audio LLMs) have advanced speech understanding, yet speaker-centric reasoning remains challenging. MSU-Bench and M3-SLU reveal gaps in speaker verification, attribution, and reasoning about who said what~\cite{msubench,kwon-etal-2025-m3slu}. VoxMem further evaluates multi-session memory for acoustic evidence, including speaker identity, and finds that current Audio LLMs retain spoken content more reliably than speaker information~\cite{xiao2026voxmem}. These findings motivate explicit speaker tools and persistent identity state.

Conversational-memory research has made substantial progress in organizing long histories. LoCoMo and LongMemEval evaluate retrieval, temporal reasoning, and knowledge updating across sessions~\cite{maharana-etal-2024-evaluating,wu-etal-2025-longmemeval}; CarMem studies preference memory for voice assistants~\cite{kirmayr-etal-2025-carmem}, while spoken and multimodal evaluations extend memory beyond text~\cite{kim-etal-2025-voice-assistant,bei-etal-2026-mem}. These benchmarks principally evaluate \emph{memory readout} from histories with available participant attribution. A shared spoken agent must also construct that history by resolving each incoming speaker and routing content to the appropriate user's memory.

Spoken dialogue systems connect identity with personalization: PAChat links speaker representations to profiles~\cite{fu-etal-2025-pachat}, AFA routes speech to per-user memories~\cite{AFA}, and EgoMem updates audiovisual memory~\cite{egomem}. Concurrent VoxPolyMem combines incremental speaker identification, gated exponential moving averages, and session-level identity reconciliation~\cite{jia2026beyond}. Yet the consequences of imperfect online speaker assignments for persistent user-memory integrity remain insufficiently understood.

A key challenge is that acoustic observations from the same speaker vary across sessions, environments, and user states. Speaker-verification research studies session variability, template aging, and adaptation~\cite{kinnunen-li-2010-overview,das-etal-2016-session,wang-etal-2016-longterm,heo-etal-2024-session}; SMIIP-TV provides recordings from 373 speakers to study such variation~\cite{qin-etal-2024-time-varying}. Linking voiceprints to writable conversational memory adds persistent consequences: false rejection can duplicate a user's profile, while false acceptance can misattribute information to another user.

We therefore argue that \emph{voiceprints are observations of identity, rather than identities themselves}. We formulate personalized spoken interaction as \emph{online speaker-memory construction}: an agent causally creates, links, and updates persistent user entities as acoustic evidence accumulates. EvoSpeakerMem routes each turn to an existing or new entity, selectively updates its acoustic state, and writes the attributed content to a replaceable textual-memory backend.

We evaluate the proposed formulation in three stages. 
First, we use MSU-Bench to examine the limitations of existing Audio LLMs in speaker-identity tasks, comparing them with an LLM equipped with a speaker-embedding tool.
Second, SMIIP-TV evaluates how evolving acoustic evidence affects speaker-memory maintenance. 
Third, LoCoMo-ID evaluates the full pipeline from online speaker routing and user-partitioned memory construction to speaker-sensitive question answering. Our contributions are summarized as follows:

\begin{itemize}
    \item We formulate \emph{online speaker-memory construction}, in which voiceprints are treated as observations of persistent user entities, and introduce EvoSpeakerMem, a training-free speech support layer for chronological \textsc{create}/\textsc{link} routing and selective speaker-state updating.

    \item We organize SMIIP-TV into static verification and chronological multi-user replay evaluations, showing how acoustic variation challenges fixed speaker representations, motivates online entity maintenance, and reveals the fragmentation--contamination trade-off.

    \item We construct LoCoMo-ID\ifx\demourl\empty\else\footnote{Interactive demo: \expandafter\url\expandafter{\demourl}}\fi, a spoken-memory benchmark with hidden speaker identities, and evaluate online speaker routing, user-partitioned memory construction, and identity-gated retrieval under clean speech, recording-condition shifts, and multi-speaker identity-confusion scenarios.
\end{itemize}

\section{Method}
\label{sec:method}

\begin{figure*}[t]
\centering
\includegraphics[width=.88\textwidth]{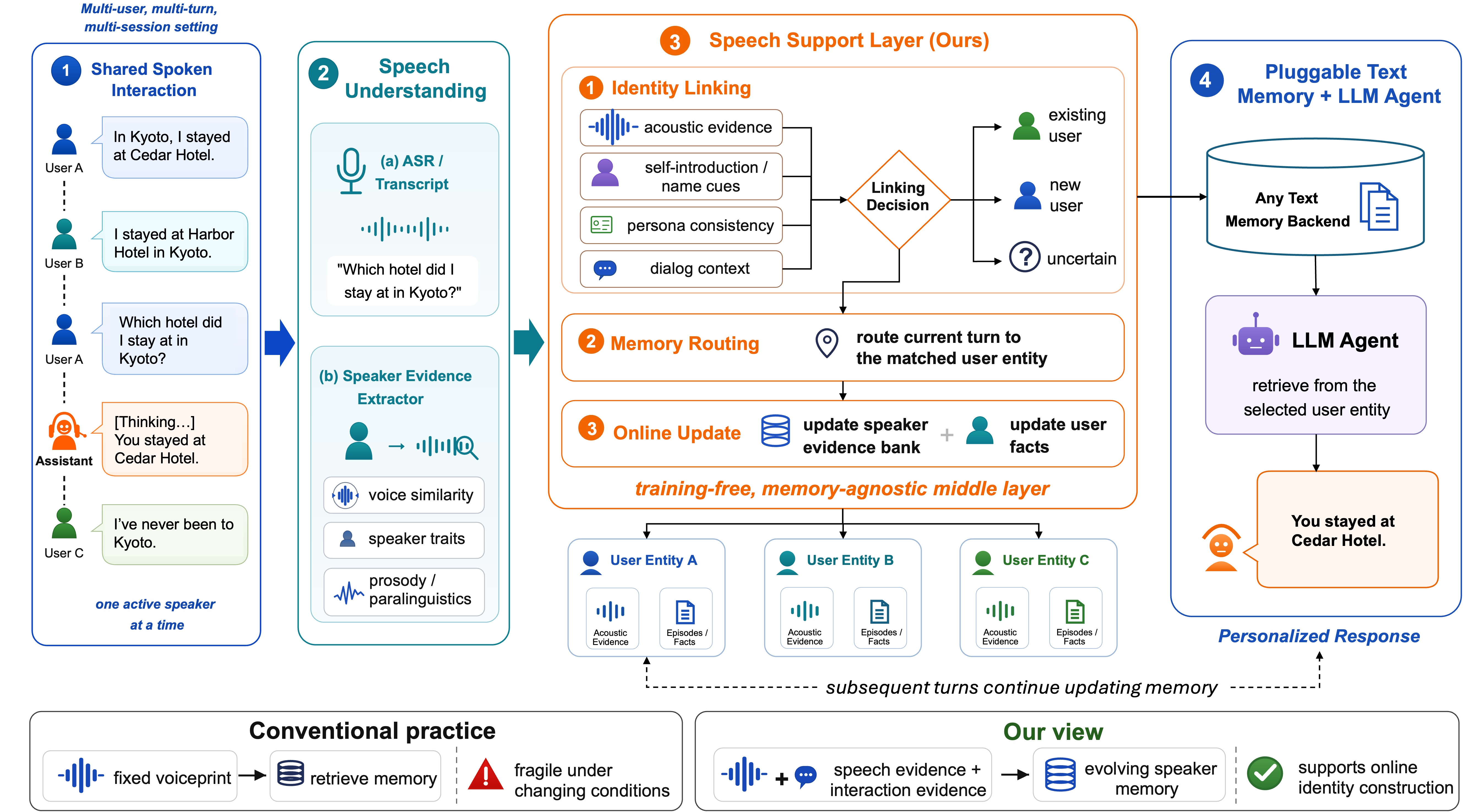}
\vspace{-2mm}\caption{Overview of EvoSpeakerMem as a speech support layer for replaceable textual-memory backends. Acoustic routing assigns each turn to a persistent user entity before its content is written to memory.}\vspace{-1mm}
\label{fig:memory-construction}
\end{figure*}

\subsection{Problem Formulation}

We consider spoken turns $\{x_t\}_{t=1}^{T}$ arriving chronologically. At time $t$, the system observes the current audio $x_t$ and its existing state $\mathcal{U}_{t-1}$, without speaker labels or access to future turns. This state contains local user entities $\mathcal{U}_{t-1}=\{u_1,\ldots,u_K\}$, each combining an acoustic representation with attributed textual memory:
\begin{equation}
    u_k=(\mathbf{c}_k,n_k,\mathcal{H}_k),
\end{equation}
where $\mathbf{c}_k$ is a speaker-embedding center, $n_k$ counts observations assigned to the entity, and $\mathcal{H}_k$ stores atomic textual facts extracted from its attributed turns. The multi-exemplar variant below replaces the center with historical embeddings.

The routing decision $r_t\in\{u_1,\ldots,u_K,\textsc{new}\}$ assigns the turn's identity and memory partition. Voiceprints are observations of persistent entities. Incorrect linking can contaminate acoustic state and factual memory; incorrect new-entity creation fragments one user's evidence across partitions.

\subsection{Online Acoustic Entity Routing}
\label{sec:routing}

A speaker encoder maps the incoming audio to a length-normalized embedding $\mathbf{e}_t$. For center-based policies, we score existing entities by cosine similarity:
\begin{equation}
    s_{t,k}=\frac{\mathbf{e}_t^\top\mathbf{c}_k}{\|\mathbf{e}_t\|\|\mathbf{c}_k\|}.
\end{equation}

The system creates a new entity initialized with $\mathbf{e}_t$ if the registry is empty or the highest entity score $s_t$ falls below $\tau_{\mathrm{merge}}$; otherwise, it links the turn to the highest-scoring entity. We set $\tau_{\mathrm{merge}}$ at the equal-error-rate (EER) operating point computed from genuine and impostor speaker-verification scores on reference voices, and keep it fixed throughout online processing.

Unrestricted updating can adapt a center to changing acoustic evidence, but an incorrect link can also make subsequent errors self-reinforcing. We therefore separate \emph{assignment} from \emph{center update}. After a turn is linked, the center changes only if $s_t\geq\tau_{\mathrm{update}}$, where $\tau_{\mathrm{update}}>\tau_{\mathrm{merge}}$ imposes a stricter requirement for modifying persistent acoustic state. An accepted update uses
\begin{equation}
    \mathbf{c}_k \leftarrow \frac{n_k\mathbf{c}_k+\mathbf{e}_t}{n_k+1}.
\end{equation}
The count $n_k$ increases on every accepted \textsc{link}, including those that leave the center unchanged. Thus a turn can contribute textual facts without necessarily changing the acoustic representation used for future matching.

We compare \textsc{Fixed}, which retains the first embedding; \textsc{Update}, which updates on every \textsc{link}; and \textsc{Gated Update}, which applies the stricter update condition. A multi-exemplar alternative stores accepted embeddings in $\mathcal{E}_k$ and scores an entity by the mean of its top $\min(m,|\mathcal{E}_k|)$ cosine similarities, for a fixed neighborhood size $m$. An embedding is appended only after the identity assignment. These policies change acoustic-state maintenance while retaining the same memory-write interface.

\subsection{Speech Support Layer and Memory Construction}
\label{sec:perturn-memory}

Figure~\ref{fig:memory-construction} illustrates the speech-to-memory pipeline. For each spoken turn, an audio LLM~\cite{xu-etal-2025-qwen3omni} generates a transcript, while a speaker-embedding model extracts acoustic identity evidence. The router (Section~\ref{sec:routing}) assigns the turn to an existing local entity or creates a new one. The transcript, assigned entity identifier, and timestamp are then passed to a replaceable text-memory backend~\cite{fang-etal-2026-lightmem}, which extracts and indexes atomic facts within the corresponding user's memory partition. Each entity thus maintains acoustic state and user-attributed textual memory, while transcript semantics do not influence acoustic routing or center updates.

At query time, read-only voice matching against the registry identifies the relevant entity and restricts retrieval to its memory partition; unmatched requests receive no personal memory. The same entity identifiers therefore govern both memory writes and reads. Appendix~\ref{app:locomo-prompts} provides transcription and routing prompts.

\section{LoCoMo-ID Benchmark Construction}
\label{sec:benchmark-construction}

LoCoMo-ID combines query adaptation, identity-consistent speech synthesis, and two stress conditions targeting acoustic variation and multi-speaker interference.

\subsection{Text Adaptation for Spoken Queries}

Following the non-commercial license, we retain LoCoMo's ten histories (5,882 turns), 1,986 QA items, and reference answers~\cite{maharana-etal-2024-evaluating}\footnote{Dataset: \url{https://huggingface.co/datasets/anonymous-user428/LoCoMo-ID}}. Rule-based query transformations replace explicit participant references with ``I'' or ``my,'' making the spoken requester's identity necessary for answering while preserving the reference answer.

We rewrite 1,371 single-person questions into first-person sensitive queries and 22 shared-experience questions into first-person-plural forms. The other 593 questions remain unchanged. Manual checks assess whether rewritten questions preserve the original meaning, participant reference, and answer. Appendix~\ref{app:synthetic-prompts} gives transformation examples.

\subsection{Voice Assignment and Speech Synthesis}

We generate candidate reference voices with Qwen3-TTS VoiceDesign\footnote{\url{https://huggingface.co/Qwen/Qwen3-TTS-12Hz-1.7B-VoiceDesign}} and retain high-quality voice reference after ECAPA-based diversity filtering. Qwen3-TTS Base\footnote{\url{https://huggingface.co/Qwen/Qwen3-TTS-12Hz-1.7B-Base}} then synthesizes the history and query audio from the assigned references~\cite{hu-etal-2026-qwen3tts}. Each conversation's two participants receive distinct voices, maintained across their history turns, sessions, and corresponding spoken queries.

Each conversation uses three voice assignments with identical dialogue text and QA content. Each conversation--realization pair initializes an independent speaker registry and textual memory. This yields 30 clean collections, 17,646 history audio turns, and 5,958 query audio items.

Quality control (QC) checks perceptual quality with UTMOS~\cite{saeki-etal-2022-utmos} and speaker consistency against a calibrated reference-similarity threshold. A failing audio clip receives up to four additional synthesis attempts; if none passes both checks, we retain the highest-UTMOS candidate. QC targets usable, identity-consistent speech rather than downstream QA. Appendix~\ref{app:synthetic-prompts} gives the voice-generation models, prompts, and QC settings.

To assess content fidelity, we transcribe history and query audio with an automatic speech recognition~(ASR) model whisper-large-v3~\cite{radford2023robust} and compute word error rate (WER) against source texts. Over 99.7\% of utterances have WER below 0.3. Very short clips can have high WER from minor ASR mismatches, such as ``bye Nate'' recognized as ``by Nate.'' Manual listening confirmed correct spoken content in these cases, supporting synthesis fidelity.

\subsection{Recording-condition Shifts}
\label{sec:spatial-construction}

To test acoustic robustness beyond the clean reference, we apply simulated room, device, and environmental variations to history audio, preserving content and speaker assignments. No quality or reference-similarity filtering follows augmentation, so difficult acoustic observations are retained. This condition simulates recording variability to mimic complex scenarios during conversations. Detailed settings are provided in Appendix~\ref{app:recording-shifts}.

\subsection{Four-Speaker Interleaved Stress Test}
\label{sec:four-speaker-construction}

We retain each original two-person story and construct a parallel story with the same conversational structure but controlled fact substitutions. The two stories share character names but use disjoint voice pairs, yielding four distinct acoustic identities with conflicting personal histories. This setting introduces deliberate identity ambiguity, testing whether the system maintains separate user memories despite overlapping names and conversational contexts.

We interleave the two stories at session boundaries while preserving the chronology of each story, requiring the system to track four identities across alternating sessions. Queries target facts from the original story, even when same-named speakers in the parallel story have conflicting information. Correct retrieval therefore depends on maintaining speaker-specific memory partitions rather than relying on textual names or shared context.


\section{Experiments}
\label{sec:experiments}

\begin{table*}[t]
\centering
\small
\setlength{\tabcolsep}{4pt}
\resizebox{.85\textwidth}{!}{%
\begin{tabular}{lllrrrr}
\hline
Condition & Acoustic evidence & Metric
& Fixed
& Update
& Gated update
& All-exemplar KNN \\
\hline

\multirow{6}{*}{Clean}
& \multirow{3}{*}{ECAPA-only}
& $F_{B^3}$ $\uparrow$
& 97.04 & 98.41 & 99.54 & 98.51 \\

&
& FNUR $\downarrow$
& 0.77 & 0.27 & 0.31 & 0.21 \\

&
& Clusters/Gold
& 3.217 & 1.767 & 1.883 & 1.583 \\

\cline{2-7}

&
\multirow{3}{*}{ECAPA + diarization}
& $F_{B^3}$ $\uparrow$
& 76.96 & 77.06 & 77.06 & 77.09 \\

&
& FNUR $\downarrow$
& 0.08 & 0.07 & 0.07 & 0.07 \\

&
& Clusters/Gold
& 1.150 & 1.117 & 1.117 & 1.117 \\

\hline

\multirow{6}{*}{\shortstack{Condition-\\shifted}}
& \multirow{3}{*}{ECAPA-only}
& $F_{B^3}$ $\uparrow$
& 74.18 & 93.54 & 92.73 & 96.62 \\

&
& FNUR $\downarrow$
& 2.00 & 0.90 & 0.94 & 0.55 \\

&
& Clusters/Gold
& 6.833 & 3.600 & 3.733 & 2.567 \\

\cline{2-7}

&
\multirow{3}{*}{ECAPA + diarization}
& $F_{B^3}$ $\uparrow$
& 81.05 & 81.23 & 81.20 & 81.04 \\

&
& FNUR $\downarrow$
& 0.15 & 0.10 & 0.10 & 0.09 \\

&
& Clusters/Gold
& 1.317 & 1.167 & 1.167 & 1.150 \\

\hline
\end{tabular}}
\caption{Online speaker-memory construction results on LoCoMo-ID clean speech and recording-shifted condition.
Acoustic evidence distinguishes ECAPA-only routing from routing augmented with session-level diarization. We report partition quality ($F_{B^3}$) and fragmentation-related indicators (FNUR and Clusters/Gold). All rates are percentages except Clusters/Gold. }
\label{tab:locomo-speaker}
\end{table*}

\subsection{Experimental Setup}

Unless otherwise stated, all online routing, causal replay, and LoCoMo-ID experiments use the SpeechBrain ECAPA-TDNN checkpoint\footnote{\url{https://huggingface.co/speechbrain/spkrec-ecapa-voxceleb}}~\cite{desplanques-etal-2020-ecapa} for acoustic embeddings. Appendices~\ref{app:redimnet2-static} and~\ref{app:redimnet2-replay} report matched-set verification and replay checks with an more advanced speaker embedding model, ReDimNet2-B6~\cite{redimnet2_2026}. History audio is processed chronologically with speaker labels hidden. 

For LoCoMo-ID, we cross four acoustic maintenance policies (\textsc{Fixed}, \textsc{Update}, \textsc{Gated Update}, and all-exemplar KNN) with two evidence configurations: per-turn ECAPA voiceprints and session-local Parakeet\cite{wang-etal-2025-ssa} diarization with ECAPA-based cross-session association. We include diarization as an alternative speaker-linking approach that groups turns by speaker within each session. However, diarization labels are session-local and cannot directly establish identity continuity across sessions, so we use ECAPA embeddings to associate these clusters with persistent user entities. Each session-local cluster is matched against the same pre-session registry snapshot before updates; the downstream memory and QA backend is otherwise shared. The transcription, memory-backend, embedding, answer-generation, judging, and decoding configurations are provided in Appendix~\ref{app:backend-config}. Our implementation is available in an anonymous code repository\footnote{\url{https://anonymous.4open.science/r/EvoSpeakerMem-4AF1}}. ECAPA embedding extraction takes 65.6\,ms per turn, excluding routing, transcription, and memory construction.

For synthesis QC, we compute the cosine similarity between each synthesized utterance and its assigned reference voice using ECAPA embeddings. We set the speaker-consistency threshold to 0.489, corresponding to the EER operating point calibrated on the VoiceDesign voice bank. This provides a verification-based criterion that balances false acceptance and false rejection on the calibration data domain. The threshold is used to assess whether synthesis preserves the intended speaker identity.


\subsection{Causal Online Evaluation}
\label{sec:online-metrics}

We evaluate online speaker-memory construction using a causal first-seen binding protocol. When a local entity is first created, it is bound to the gold speaker of that turn only for post-hoc scoring. The binding is immutable and never affects routing, updating, or memory construction; a later assignment is correct only when it matches this binding. This avoids future-aware cluster matching, which could relabel entities after observing the full sequence and hide online routing errors.

Each turn is categorized into one of five mutually exclusive routing outcomes: new-speaker creation, new-speaker linking, known-speaker creation, correct known-speaker linking, and wrong-target linking. These outcomes capture two complementary failure modes of online entity maintenance: fragmentation, where a known speaker is assigned to a new entity, and contamination, where different speakers are merged into the same entity.

B$^3$ F1~\cite{bagga-baldwin-1998-entity-based} penalizes cross-speaker merging and within-speaker fragmentation. Clusters/Gold is the predicted-to-gold entity count ratio (ideal: 1). \textsc{Create} precision ($P_{\mathrm{create}}$) is correct creations/all \textsc{creates}; recall ($R_{\mathrm{create}}$) uses first-seen speaker turns as denominator. \textsc{Link} precision ($P_{\mathrm{link}}$) is correct links/all \textsc{links}. Known-user recall ($R_{\mathrm{known}}$) and false-new user rate (FNUR) divide correct links and erroneous creations, respectively, by returning-speaker turns. Online ID accuracy is (correct creations + correct links)/evaluated turns, excluding missing predictions. Appendix~\ref{app:online-metrics} gives the equivalent count-based formulas.

\subsection{MSU-Bench: Motivating Explicit Speaker State}
\label{sec:msu}

We use the speaker retrieval (SR) and verification (SV) tasks in MSU-Bench to examine the value of explicit acoustic identity tools and persistent speaker state. Using annotation-provided temporal boundaries but no gold speaker identities, an Audio LLM invokes an ECAPA-based speaker tool for each segment in chronological order. The tool creates or links local speaker entities, maintains a persistent speaker registry, and provides deterministic query-time readout. This configuration achieves 96.61\% SR and 79.35\% SV accuracy, exceeding the reported end-to-end Audio LLM results. Detailed results appear in Appendix~\ref{app:msu-details}.

\subsection{Time-Varying Speaker Evidence on SMIIP-TV}
\label{sec:smiip-tv}

\subsubsection{Longitudinal Verification Setup}

SMIIP-TV~\cite{qin-etal-2024-time-varying}\footnote{\url{https://www.openslr.org/156}} contains repeated recordings from 373 speakers over 90 days, with an approximately one-year follow-up for 118 matched speakers. Its text-independent utterances cover variations in recording environment, speaker state, and time of day.

We use each speaker's earliest utterance as enrollment and evaluate a frozen ECAPA representation across increasing time gaps. Using impostor scores from probes within three days of enrollment, we calibrate a threshold at approximately 2\% false acceptance ($\tau_{2\%}=0.4751$) and freeze it for subsequent analyses. We report EER and \emph{Below-$\tau$}, the proportion of genuine probes falling below this threshold. Complete protocol details are provided in Appendix~\ref{app:smiip-static-protocol}.

\subsubsection{Fixed-Template Degradation}

\begin{table}[t]
\centering
\setlength{\tabcolsep}{3pt}
\resizebox{.95\columnwidth}{!}{
\begin{tabular}{@{}lrrrr@{}}
\hline
Gap (days) & EER & Below-$\tau$ & Imp. accept & Gen. sim. \\
\hline
0--1 days & 9.7 & 29.8 & 2.0 & 0.538 \\
60--100 days & 11.2 & 40.7 & 1.8 & 0.495 \\
365--450 days & 14.0 & 55.6 & 1.4 & 0.446 \\
\hline
\end{tabular}
}
\caption{Fixed-template verification across SMIIP-TV time gaps. Error rates are percentages. ``Imp. accept'' is the proportion of impostor trials exceeding the frozen threshold; ``Gen. sim.'' is mean genuine cosine similarity.}
\label{tab:smiip-static}
\end{table}

Table~\ref{tab:smiip-static} shows that fixed-template reliability deteriorates with increasing enrollment--probe gaps. Below-$\tau$ rises from 29.8\% within one day to 55.6\% after approximately one year, while EER increases from 9.7\% to 14.0\%. The trend also persists in an observational subset restricted to bedroom/dormitory recordings, normal speaker state, and normal noise: Below-$\tau$ rises from 16.42\% to 29.71\% between the 0--1- and 60--100-day intervals (Appendix~\ref{app:smiip-context}).

A stronger encoder mitigates but does not eliminate this effect. On matched SMIIP-TV utterances, ReDimNet2-B6 reduces absolute verification errors, yet its Below-$\tau$ rate still increases from 5.19\% to 12.96\% across the same time gaps. These findings indicate that stronger acoustic representations alone do not make a fixed enrollment voiceprint a reliable long-term identity key. Detailed encoder comparisons appear in Appendix~\ref{app:redimnet2-static}.

\subsubsection{Causal Speaker Maintenance}

To examine how acoustic variability affects persistent identity construction, we organize SMIIP-TV recordings into causal replay streams containing two to four speakers. We evaluate six conditions involving temporal, environmental, and speaker-state variation, with five seeds per condition. Each turn is processed chronologically without future speaker information. We compare six training-free representation policies, ranging from fixed and updated single templates to all-exemplar KNN (Appendices~\ref{app:smiip-spans} and~\ref{app:smiip-online}).

Across the six large-span conditions, online mean updating improves macro-averaged B$^3$ F1 from 46.82\% (First exemplar) to 55.99\%, while All-exemplar KNN achieves 65.31\%. However, preserving more historical evidence does not guarantee correct identity assignment: All-exemplar KNN achieves lower \textsc{link} precision than First exemplar (52.56\% vs.\ 66.25\%). These results show that adaptive representations can mitigate fragmentation, but also expose a trade-off with incorrect linking. We examine the downstream consequences of this trade-off through memory construction and identity-gated QA on LoCoMo-ID (Section~\ref{sec:locomo}).

\subsection{Online Speaker-Memory Evaluation on LoCoMo-ID}
\label{sec:locomo}

\subsubsection{History Routing and Entity Structure}

Table~\ref{tab:locomo-speaker} compares four maintenance policies in two evidence modes across 30 clean collections and 30 condition-shifted collections. We fix $\tau_{\mathrm{merge}}$ for routing and private-query lookup, and a stricter threshold $\tau_\mathrm{update}$ for gated updates. In this paper, we calibrate $\tau_{\mathrm{merge}}$ on the VoiceDesign voice bank to adapt the speaker-recognition decision boundary to the score distribution of the target voice domain. Its EER operating point, 0.489, provides a domain-calibrated matching criterion shared by all maintenance policies. For gated updates, we use the threshold $\tau_{\mathrm{update}}=0.58$, supported by the clean-condition ablation in Appendix~\ref{app:locomo-sensitivity}: performance approaches a plateau at this value, and further increases offer only marginal additional benefits. Parakeet supplies within-session labels and ECAPA performs cross-session association, whereas the global-voiceprint configuration routes individual turns without the session label grouping.

Table~\ref{tab:locomo-speaker} shows that, across both evidence modes and both speech conditions, Update, Gated Update, and All-exemplar KNN reduce FNUR and Clusters/Gold and generally improve B$^3$ F1 relative to Fixed. The largest gains occur under ECAPA-only routing with shifted conditions, where B$^3$ F1 rises from 74.18\% for Fixed to 92.73--96.62\% for the non-fixed policies. Overall, maintaining speaker evidence beyond the initial voiceprint improves entity structure. Full causal CREATE/LINK diagnostics appear in Appendix~\ref{app:locomo-private}.

\subsubsection{All-Memory Retrieval and QA}

\begin{table}[t]
\centering
\resizebox{\columnwidth}{!}{%
\begin{tabular}{lrr}
\hline
Acoustic evidence / policy & Clean & Condition-shifted\\
\hline
ECAPA / Fixed & 72.31 & 69.27 \\
ECAPA / Update & 73.30 & 72.55 \\
ECAPA / Gated update & 73.38 & 72.04 \\
ECAPA / All-exemplar KNN & 73.38 & 72.79 \\
Parakeet+ECAPA / Fixed & 73.43 & 72.96 \\
Parakeet+ECAPA / Update & 73.50 & 73.16 \\
Parakeet+ECAPA / Gated update & \textbf{73.52} & \textbf{73.40} \\
Parakeet+ECAPA / All-exemplar KNN & 73.33 & 73.18 \\
\hline
\end{tabular}}
\caption{Matched all-memory QA (\%) for eight policies at merge threshold 0.489. Results are micro-pooled judge-correct answers over speaker-sensitive queries without query-speaker filtering or privacy denial.}
\label{tab:locomo-all-qa}
\end{table}

All-memory QA transcribes each query and retrieves up to 60 items across the collection without identifying the requester or filtering by speaker. Table~\ref{tab:locomo-all-qa} evaluates the same 4,113 sensitive queries
used below, enabling comparison of unrestricted and identity-gated retrieval. ECAPA gated updating reaches 73.38\% versus 72.31\% for Fixed on clean speech, and 72.04\% versus 69.27\% under shifted-condition.
Because all-memory retrieval does not depend on assigning the query to a speaker-specific entity, differences among speaker-memory policies are smaller here than under identity-gated retrieval, particularly on clean
speech. Nevertheless, the gains under shifted-condition show that updating speaker evidence can still improve the quality and coverage of the retrieved memory, even when access is unrestricted. Full-set results
on all 5,958 queries appear in Appendix~\ref{app:locomo-all-memory-full}.

\subsubsection{Identity-Gated Retrieval}

\begin{table}[t]
\centering
\resizebox{\columnwidth}{!}{
\begin{tabular}{llrrr}
\hline
Condition & ECAPA policy & C/W/D & QA & QA$\mid$Correct \\
\hline
Clean & Fixed & 4084/0/29 & 71.41 & 71.91 \\
 & Update & 3993/115/5 & 73.43 & 75.63 \\
 & Gated update & 4110/0/3 & \textbf{75.27} & 75.33 \\
 & KNN & 3993/116/4 & 73.64 & \textbf{75.86} \\
\hline
Condition-shifted& Fixed & 4079/1/33 & 60.15 & 60.65 \\
 & Update & 3993/115/5 & 70.24 & 72.35 \\
 & Gated update & 4109/0/4 & 71.77 & 71.84 \\
 & KNN & 3993/116/4 & \textbf{72.40} & \textbf{74.58} \\
\hline
\end{tabular}
}
\caption{Top-1 private-access results on 4,113 speaker-sensitive query audios under clean and shifted condition. C/W/D denotes correct, wrong, and denied authorizations; QA requires correct authorization and a judge-correct answer (\%). KNN denotes All-exemplar KNN.}
\label{tab:locomo-sensitive-qa}
\end{table}

\begin{table*}[t]
\centering
\small
\setlength{\tabcolsep}{4pt}
\resizebox{.9\textwidth}{!}{%
\begin{tabular}{lrrrrr}
\hline
\multicolumn{1}{c}{\begin{tabular}[c]{@{}c@{}}Acoustic evidence / policy\end{tabular}}
& \multicolumn{1}{c}{\begin{tabular}[c]{@{}c@{}}Online ID\end{tabular}}
& \multicolumn{1}{c}{\begin{tabular}[c]{@{}c@{}}$F_{B^3}$\end{tabular}}
& \multicolumn{1}{c}{\begin{tabular}[c]{@{}c@{}}C / W / D\end{tabular}}
& \multicolumn{1}{c}{\begin{tabular}[c]{@{}c@{}}Targeted QA\\($n=19$)\end{tabular}}
& \multicolumn{1}{c}{\begin{tabular}[c]{@{}c@{}}Full QA\\($n=1{,}371$)\end{tabular}} \\
\hline
ECAPA / Fixed & \textbf{98.84} & 96.54 & 1342 / 18 / 11 & \textbf{78.95} & 67.83 \\
ECAPA / Update & 93.35 & 95.45 & 1335 / 32 / 4 & \textbf{78.95} & 72.06 \\
ECAPA / Gated update & 96.25 & \textbf{97.77} & 1368 / 0 / 3 & \textbf{78.95} & \textbf{73.60} \\
ECAPA / All-exemplar KNN & 94.85 & 97.20 & 1335 / 33 / 3 & \textbf{78.95} & 71.99 \\
\hline
Parakeet+ECAPA / Fixed & 74.03 & 72.86 & 1335 / 32 / 4 & 68.42 & 71.48 \\
Parakeet+ECAPA / Update & 71.96 & 72.57 & 1333 / 32 / 6 & 73.68 & 72.28 \\
Parakeet+ECAPA / Gated update & 73.87 & 73.10 & 1334 / 32 / 5 & 68.42 & 72.21 \\
Parakeet+ECAPA / All-exemplar KNN & 71.93 & 72.60 & 1335 / 32 / 4 & 73.68 & 72.36 \\
\hline
\end{tabular}}
\caption{Four-speaker interleaved stress test: ten clean collections, 11,764 history turns, and 40 gold speakers. Online ID excludes missing predictions. C/W/D counts correct/wrong/denied authorizations on 1,371 queries. Both QA columns require correct authorization and a judge-correct answer; targeted QA uses 19 gated items among 20 diagnostics. Rates are percentages. This differs from the standard 4,113-query condition.}\vspace{-3mm}
\label{tab:locomo-interleaved}
\end{table*}

For the 4,113 first-person, speaker-sensitive query audios per condition, ECAPA performs read-only lookup against the registry constructed from that condition's history turns. Top-1 authorizes the highest-scoring entity only at similarity $\tau_{\mathrm{merge}}=0.489$ or above; otherwise no personal memory is returned. The authorized identifier is a hard retrieval filter. Queries never create entities, append exemplars, update centers, or write to LightMem. Gold requester identities are used solely for post-hoc scoring. A no-match query is an upstream refusal; a downstream answer refusal after authorization is not counted as an authorization denial.

We evaluate Top-1 authorization against the immutable entity-origin binding used for history-routing scoring. A query is \emph{correctly authorized} when the selected entity originates from the gold requester, \emph{wrongly authorized} when it originates from another user, and \emph{denied} when no entity reaches the threshold. Let $C$, $W$, and $D$ be the mutually exclusive counts of these outcomes and $N_q=C+W+D$.

Let $A_q\in\{\mathrm{correct},\mathrm{wrong},\mathrm{denied}\}$ denote the authorization outcome for query $q$, and let $J_q\in\{0,1\}$ denote the binary LoCoMo judgment of answer correctness. To couple access correctness with answer utility, we define end-to-end identity-gated QA as
\begin{equation}
\mathrm{QA}
=\frac{1}{N_q}\sum_{q=1}^{N_q}
\mathbf{1}[A_q=\mathrm{correct}]J_q.
\end{equation}
Denials, wrong authorizations, and incorrect or unparseable answers
therefore receive zero credit. We additionally report
\begin{equation}
\mathrm{QA\mid Correct}
=\frac{1}{C}\sum_{q=1}^{N_q}
\mathbf{1}[A_q=\mathrm{correct}]J_q,
\end{equation}
which isolates memory retrieval and answer generation among correctly
authorized queries.

Table~\ref{tab:locomo-sensitive-qa} reports identity-gated QA for ECAPA-only routing, with corresponding diarization-augmented results in Appendix~\ref{app:locomo-private}. With ECAPA-only routing, QA rises from 71.41\% for Fixed to 73.43--75.27\% on clean audio and from 60.15\% to 70.24--72.40\% under shifted condition. The same pattern holds for Parakeet+ECAPA: QA rises from 68.08\% to 68.10--68.25\% on clean audio and from 63.36\% to 67.20--67.76\% under shifted condition. These results show that retaining or updating speaker evidence beyond the initial voiceprint improves downstream personalized QA, with the largest gains under shifted condition. 

This reveals a trade-off between answer utility and identity integrity. Under shifted conditions, All-exemplar KNN achieves slightly higher identity-gated QA than Gated Update (72.40\% vs.\ 71.77\%), but incurs 116 wrong authorizations versus none for Gated Update, highlighting the importance of preventing cross-user memory access. 

Identity filtering can restrict evidence coverage: Fixed's shifted condition all-memory QA is 69.27\%, versus 60.15\% strict QA. Appendix~\ref{app:locomo-private} discusses fragmentation and this utility trade-off.

With merge/lookup threshold fixed at 0.489, a clean update-threshold sweep tests sensitivity of the policy comparison: online accuracy and B$^3$ F1 are nearly flat from 0.58 to 0.66, and wrong authorizations remain zero through 0.70, versus 115 at 0.50 and 32 at 0.54. Duration analysis finds lower gated-update accuracy for 1--$<2$\,s turns: 98.17\% clean and 77.38\% shifted condition, versus 99.74\% and 99.10\% for $\geq 2$\,s. Appendix~\ref{app:locomo-sensitivity} provides both diagnostics.

\subsubsection{Four-Speaker Interleaved Stress Test}
\label{sec:four-speaker-results}

Compared with the standard two-speaker setting, the four-speaker interleaved stress test doubles the speakers per conversation and adds a fact-conflicting parallel story with shared character names, increasing identity and memory interference (Table~\ref{tab:locomo-interleaved}). Descriptively, ECAPA Gated Update attains 97.77\% B$^3$ F1 and 73.60\% full identity-gated QA in the four-speaker setting, compared with 99.54\% and 75.27\% in the standard clean two-speaker setting. Crucially, updating speaker evidence remains beneficial in the more complex setting: ECAPA Update, Gated Update, and All-exemplar KNN achieve 72.06\%, 73.60\%, and 71.99\% full QA, respectively, versus 67.83\% for Fixed; Gated Update also has the highest four-speaker B$^3$ F1 (97.77\%). The same QA advantage over Fixed holds with Parakeet+ECAPA (72.21--72.36\% versus 71.48\%). Appendix~\ref{app:locomo-four-speaker} provides detailed results.

\section{Conclusion}

We formulate \emph{online speaker-memory construction}, placing speaker attribution in the memory write path. Longitudinal speech challenges fixed templates; clean speech still exposes fragmentation. Across clean and condition-shifted LoCoMo-ID experiments, Update, Gated Update, and All-exemplar KNN improve identity-gated QA over Fixed in both acoustic evidence modes. This advantage persists in the more challenging four-speaker interleaved test: with ECAPA-only routing, the three non-fixed policies achieve 71.99--73.60\% full QA versus 67.83\% for Fixed, with Gated Update attaining the highest score. These findings support maintaining speaker evidence beyond the initial voiceprint when constructing and querying personalized memory.

\vfill
\pagebreak

\section*{Limitations}

Our evaluation has several limitations. LoCoMo-ID uses synthetic voices; shifted-condition simulates session-dependent rooms, microphones, device chains, and noise rather than naturally recorded, overlapping multi-user conversations. The standard clean and shifted-condition collections each contain two participants. Our four-speaker stress test interleaves two synthetic, fact-conflicting stories at session boundaries, preserving within-story chronology; it does not represent naturally recorded, overlapping four-party dialogue. Its full-query evaluation covers only the ten clean collections. The two-to-four-speaker SMIIP-TV streams evaluate entity maintenance over repeated recordings rather than dialogue-memory QA. The Parakeet comparison combines session-local diarization with the same ECAPA cross-session registry, not an independently evaluated published system. We vary only the center-update threshold on clean speech: merge/lookup threshold 0.489 has no independent sensitivity sweep, and the operational encoder and backend remain fixed. Incorrectly written memories are not automatically corrected.

Although multi-exemplar KNN is now evaluated through the LoCoMo-ID write/read pipeline, it stores every accepted voiceprint and does not bound memory growth. The condition-shifted evaluation shows improved clustering but wrong-user authorization, so a bounded, contamination-resistant exemplar selection policy remains future work. The post-hoc duration bins (clean 59/17,643 and condition-shifted 13/17,644 sub-second evaluated turns for global ECAPA) contain few short turns, so they cannot establish robust duration effects or the influence of encoder padding; no duration-controlled intervention or duration-stratified private QA is available. Identity-gated retrieval is a simulated access protocol, not a formal privacy or authentication guarantee.

\section*{Ethical Considerations}

Speaker embeddings and persistent conversational memory can reveal sensitive information. A local anonymous identifier does not make a person's voice anonymous, and long-term voice linkage could be misused for surveillance or identification beyond the intended interaction context. Raw audio, embeddings, transcripts, and extracted memories should therefore be collected and processed only with appropriate consent and access controls, retained for no longer than necessary, and protected against unauthorized reuse. Synthetic voices and derived conversational data should likewise be used in accordance with the licenses and intended uses of their source assets. The synthetic voice bank is constructed from anonymous generated references and is not intended to imitate or identify real individuals.

Identity-routing errors also create direct privacy risks: a false merge may expose one user's information to another, while fragmented profiles can make correction and deletion incomplete. The observed wrong authorizations under unrestricted center updating, multi-exemplar scoring, and session-log-based grouping illustrate how plausible matches can expose another user's memory under the simulated protocol. A deployed system should provide transparent enrollment and identity-management mechanisms, support user inspection, correction, deletion, and withdrawal, and avoid using the proposed router as a security-critical authentication mechanism. Our experiments study the technical problem of identity-aware memory construction and should be understood as a research prototype rather than a deployment-ready biometric system.

\bibliography{custom,related_work,anthology-overleaf-1,anthology-overleaf-2}


\appendix
\raggedbottom

\section{Detailed Online Routing Metrics}
\label{app:online-metrics}

Following the causal first-seen binding protocol in Section~\ref{sec:online-metrics}, let $a$, $b$, $c$, $d$, and $e$ denote new speaker--\textsc{create}, new speaker--\textsc{link}, known speaker--\textsc{create}, known speaker--correct \textsc{link}, and known speaker--wrong-target \textsc{link}, respectively, with $N=a+b+c+d+e$. A speaker is considered new if it has not appeared earlier in the causal stream, and a \textsc{link} is correct only when the selected entity matches the immutable first-creation binding.

We report detailed routing statistics, including \textsc{create} precision~($P_{\mathrm{create}}$), \textsc{create} recall~($R_{\mathrm{create}}$), \textsc{link} precision~($P_{\mathrm{link}}$), and known-user resolution recall~($R_{\mathrm{known}}$). FNUR is reported in the main paper as a causal fragmentation indicator.
\begin{equation}
\begin{aligned}
P_{\mathrm{create}}
  &= \frac{a}{a+c},
& R_{\mathrm{create}}
  &= \frac{a}{a+b},\\
P_{\mathrm{link}}
  &= \frac{d}{b+d+e},
& R_{\mathrm{known}}
  &= \frac{d}{c+d+e},\\
\mathrm{FNUR}
  &= \frac{c}{c+d+e}.
\end{aligned}
\end{equation}

Known-user resolution recall measures correct resolution of previously observed speakers while penalizing erroneous \textsc{create} and wrong-target \textsc{link} decisions. FNUR captures the fraction of returning-speaker turns incorrectly assigned to newly created entities.
Online ID accuracy is $(a+d)/N$; turns without a prediction are excluded from $N$.

\section{Additional MSU-Bench Results}
\label{app:msu-details}

\begin{table}[t]
\centering
\small
\setlength{\tabcolsep}{3pt}
\begin{tabular}{@{}p{0.70\columnwidth}rr@{}}
\hline
Model & SR & SV \\
\hline
Qwen2.5-Omni~\cite{xu-etal-2025-qwen25omni} & 19.00 & 24.00 \\
AudioFlamingo-3~\cite{goel-etal-2025-audioflamingo3} & 23.00 & 29.00 \\
Qwen3-Omni~\cite{xu-etal-2025-qwen3omni} & 49.00 & 40.00 \\
Kimi-Audio~\cite{kimi-team-2025-kimiaudio} & 46.00 & 24.00 \\
StepAudio2~\cite{wu-etal-2025-stepaudio2} & 34.00 & 43.00 \\
MiMoAudio~\cite{xiaomi-2025-mimoaudio} & 38.00 & 48.00 \\
Gemini-2.5-Flash~\cite{comanici-etal-2025-gemini25} & 66.00 & 68.00 \\
Gemini-2.5-Pro~\cite{comanici-etal-2025-gemini25} & 67.00 & 68.00 \\
Gemini-3-Flash~\cite{google-deepmind-2025-gemini3flash} & 69.00 & 75.00 \\
\hline
Unified speaker state & \textbf{96.61} & \textbf{79.35} \\
\hline
\end{tabular}
\caption{MSU-Bench speaker retrieval (SR) and verification (SV) accuracy (\%). Unified speaker state uses deterministic tool readout.}\vspace{-4mm}
\label{tab:msu}
\end{table}

Table~\ref{tab:msu} compares this tool-augmented persistent speaker state with reported end-to-end Audio LLM results on MSU-Bench. It achieves 96.61\% on SR and 79.35\% on SV, exceeding every listed end-to-end model on the respective task. The comparison shows that an Audio LLM can attain strong speaker-memory performance by orchestrating an explicit lightweight acoustic identity module and persistent state, rather than relying solely on direct answer generation.

\section{Additional SMIIP-TV Analyses}
\label{app:smiip-details}

\begin{table*}[t]
\centering
\small
\setlength{\tabcolsep}{4pt}
\resizebox{.6\textwidth}{!}{%
\begin{tabular}{lrrrrrr}
\hline
& \multicolumn{2}{c}{0--1 days}
& \multicolumn{2}{c}{60--100 days}
& \multicolumn{2}{c}{365--450 days} \\
\cline{2-7}
Encoder
& EER & Below-$\tau$
& EER & Below-$\tau$
& EER & Below-$\tau$ \\
\hline
ECAPA-TDNN
& 9.71 & 29.82
& 11.18 & 40.70
& 14.09 & 55.54 \\
ReDimNet2-B6
& 3.34 & 5.19
& 4.82 & 8.78
& 6.35 & 12.96 \\
\hline
\end{tabular}%
}
\caption{Matched-set fixed-template verification at representative
enrollment--probe gaps. EER and Below-$\tau$ are percentages.
The comparison uses the 205,765 text-independent utterances successfully
processed by both encoders, with identical enrollment utterances, genuine
probes, impostor trials, and time buckets. Each encoder uses its own
threshold frozen after near-enrollment calibration at target FAR=2\%
($\tau_{\mathrm{ECAPA}}=0.4752$;
$\tau_{\mathrm{ReDimNet2}}=0.6131$), so threshold values are not compared
directly. ReDimNet2-B6 lowers absolute error, but both encoders exhibit
higher EER and a larger below-threshold genuine rate at longer gaps.}
\label{tab:app-redimnet2-static}
\end{table*}

\begin{figure*}[t]
    \centering
    \includegraphics[width=\textwidth]{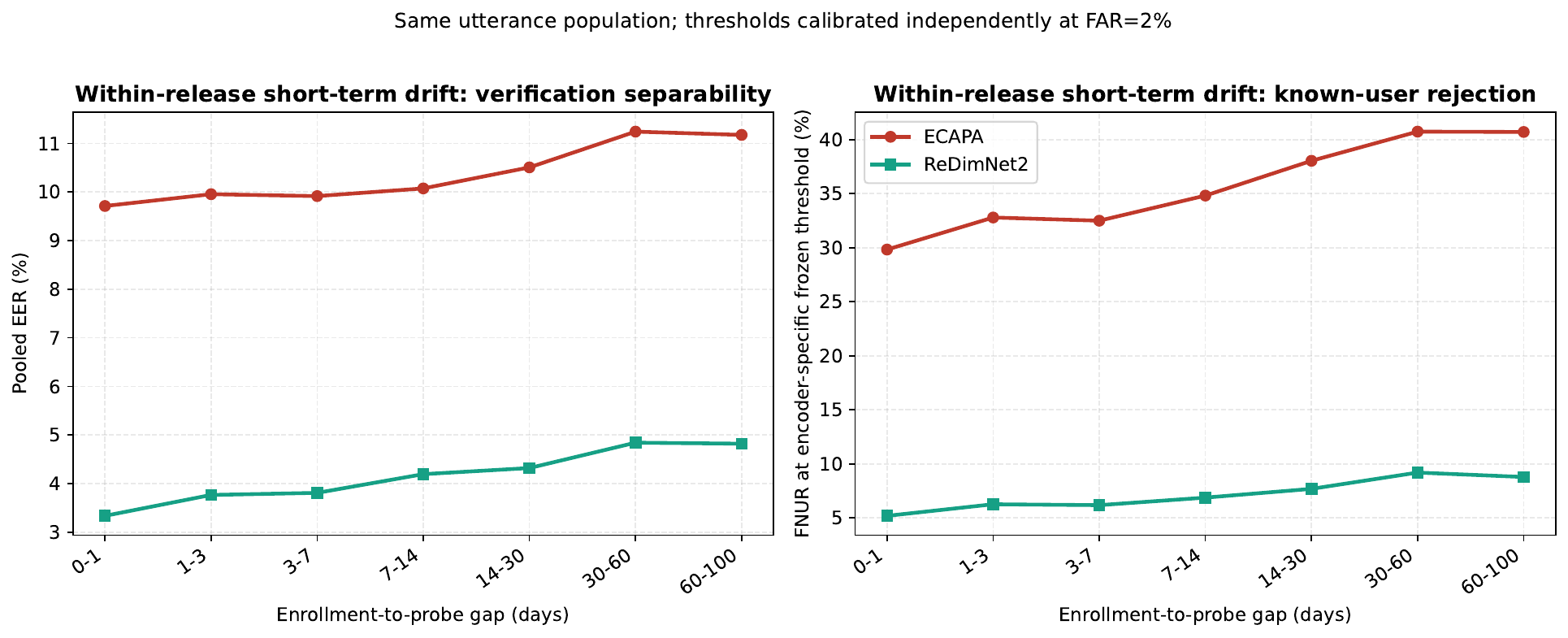}
    \caption{Matched-set short-term fixed-template verification for ECAPA
and ReDimNet2-B6. Both encoders use identical utterances and time
buckets, while each uses an encoder-specific threshold frozen after
near-enrollment FAR=2\% calibration. ReDimNet2-B6 lowers absolute EER
and below-threshold genuine rate, but both measures increase as the
enrollment--probe gap grows.}
    \label{fig:comp-smiip-shortterm}
\end{figure*}

\begin{figure*}[t]
    \centering
    \includegraphics[width=\textwidth]{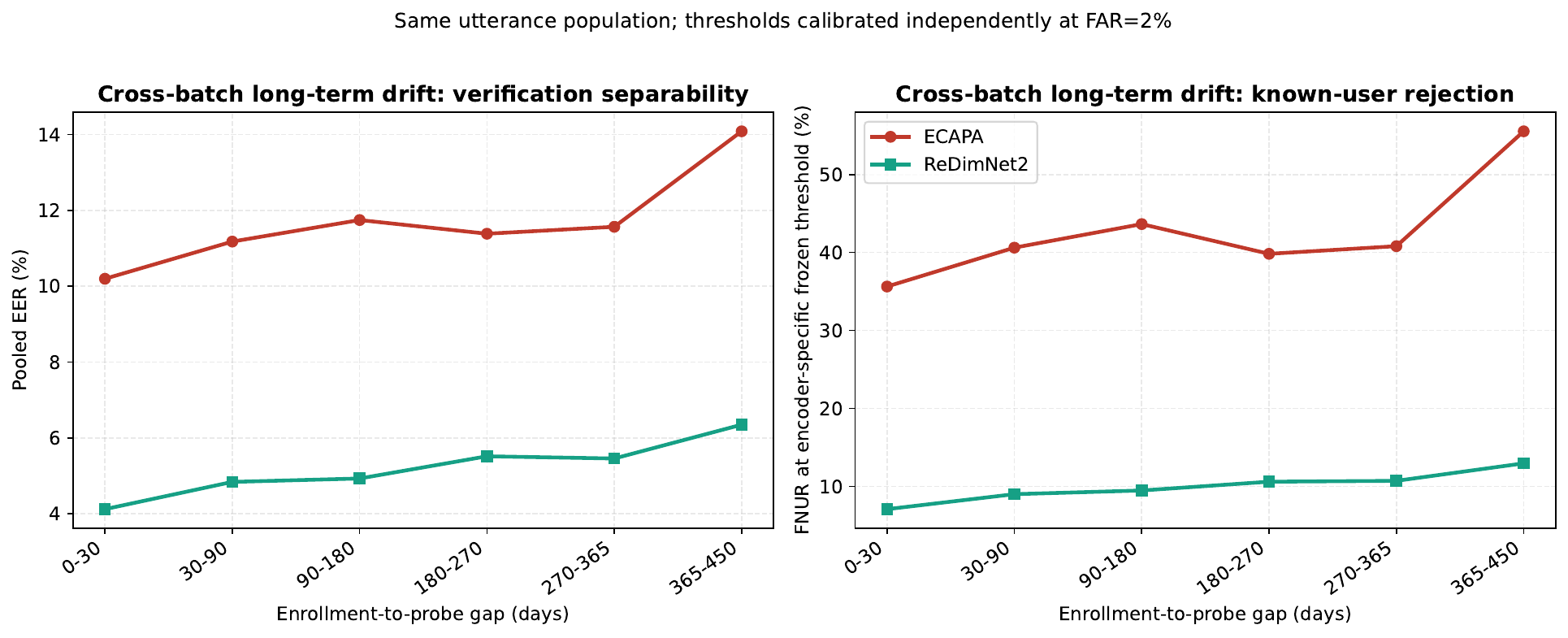}
    \caption{Matched-set long-term fixed-template verification for ECAPA
and ReDimNet2-B6, including the approximately one-year follow-up.
Although ReDimNet2-B6 substantially improves absolute separability, its
EER and below-threshold genuine rate still worsen at longer
enrollment--probe gaps.}
    \label{fig:comp-smiip-longterm}
\end{figure*}

\begin{table*}[t]
\centering
\small
\setlength{\tabcolsep}{4pt}
\resizebox{.7\textwidth}{!}{%
\begin{tabular}{lrrrrrrr}
\hline
& \multicolumn{3}{c}{ECAPA-TDNN}
& \multicolumn{3}{c}{ReDimNet2-B6} \\
\cline{2-7}
Large-span condition
& $F_{B^3}$ $\uparrow$ & FNUR $\downarrow$ & Frag. $\downarrow$
& $F_{B^3}$ $\uparrow$ & FNUR $\downarrow$ & Frag. $\downarrow$ \\
\hline
Intra-day TOD
& 63.08 & 18.09 & 25.52
& 62.91 & 6.75 & 12.33 \\
Inter-day
& 66.58 & 1.21 & 12.20
& 70.56 & 0.22 & 3.08 \\
Long-term cross-batch
& 67.60 & 3.60 & 20.79
& 81.49 & 0.91 & 4.51 \\
Environment
& 67.05 & 5.00 & 23.53
& 73.43 & 1.16 & 7.84 \\
Speaker state
& 58.88 & 6.78 & 27.63
& 66.90 & 1.14 & 11.45 \\
Long-term $\times$ hist.-env.
& 69.54 & 4.10 & 21.68
& 86.78 & 1.02 & 4.29 \\
\hline
Mean
& 65.45 & --- & ---
& 73.68 & --- & --- \\
\hline
\end{tabular}%
}
\caption{Matched-session large-span causal replay for the
all-exemplar KNN policy. ECAPA-TDNN and ReDimNet2-B6 replay identical
common-utterance session streams under encoder-specific thresholds
calibrated with the same near-enrollment FAR=2\% protocol. Frag. denotes
speaker fragmentation (\%). ReDimNet2-B6 improves absolute online entity
maintenance in most conditions, particularly for long-term and
context-variable replay; the near-equal intra-day B$^3$ F1 should be
interpreted together with its lower FNUR and fragmentation.}
\label{tab:app-redimnet2-replay}
\end{table*}

\begin{figure*}[t]
    \centering
    \includegraphics[width=\textwidth]{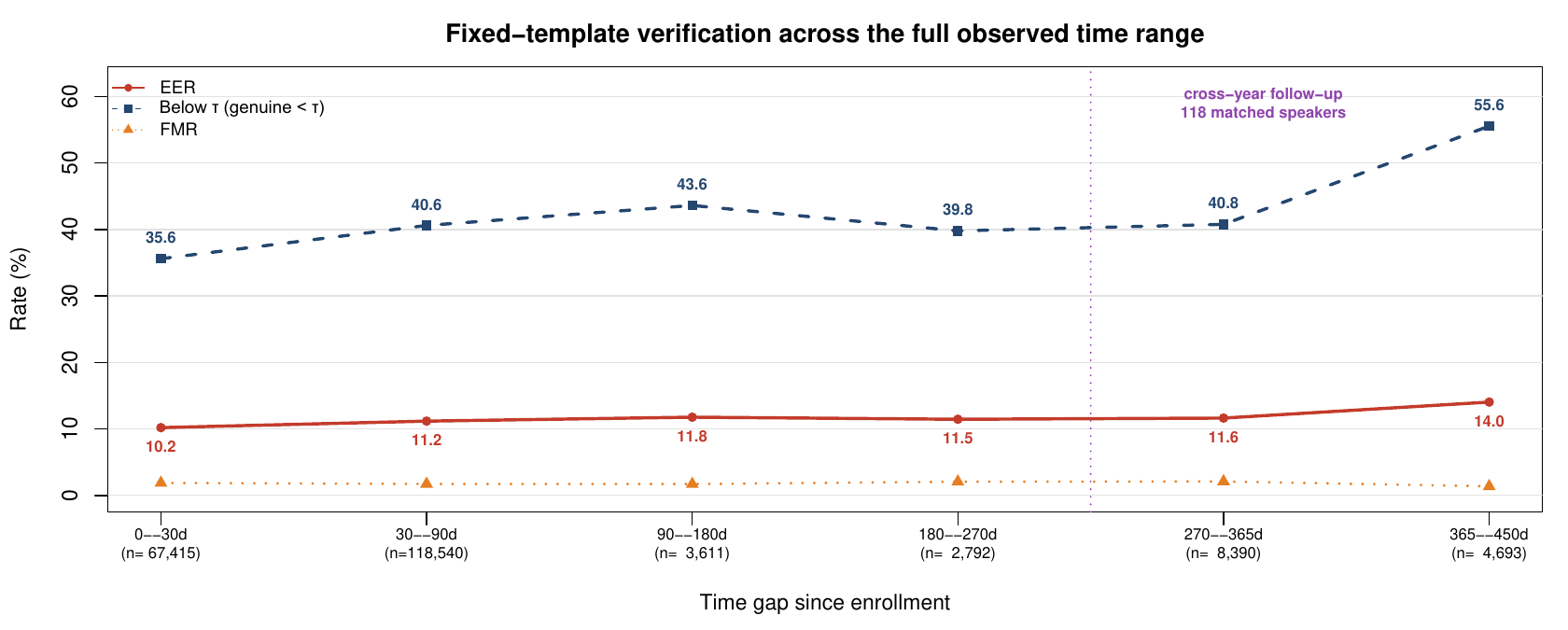}
    \caption{Fixed-template verification over the full observed time range. The same low-drift-calibrated threshold is frozen for all buckets; the final bucket uses the approximately one-year follow-up collection.}
    \label{fig:app-smiip-longterm}
\end{figure*}

\begin{figure*}[t]
    \centering
    \includegraphics[width=\textwidth]{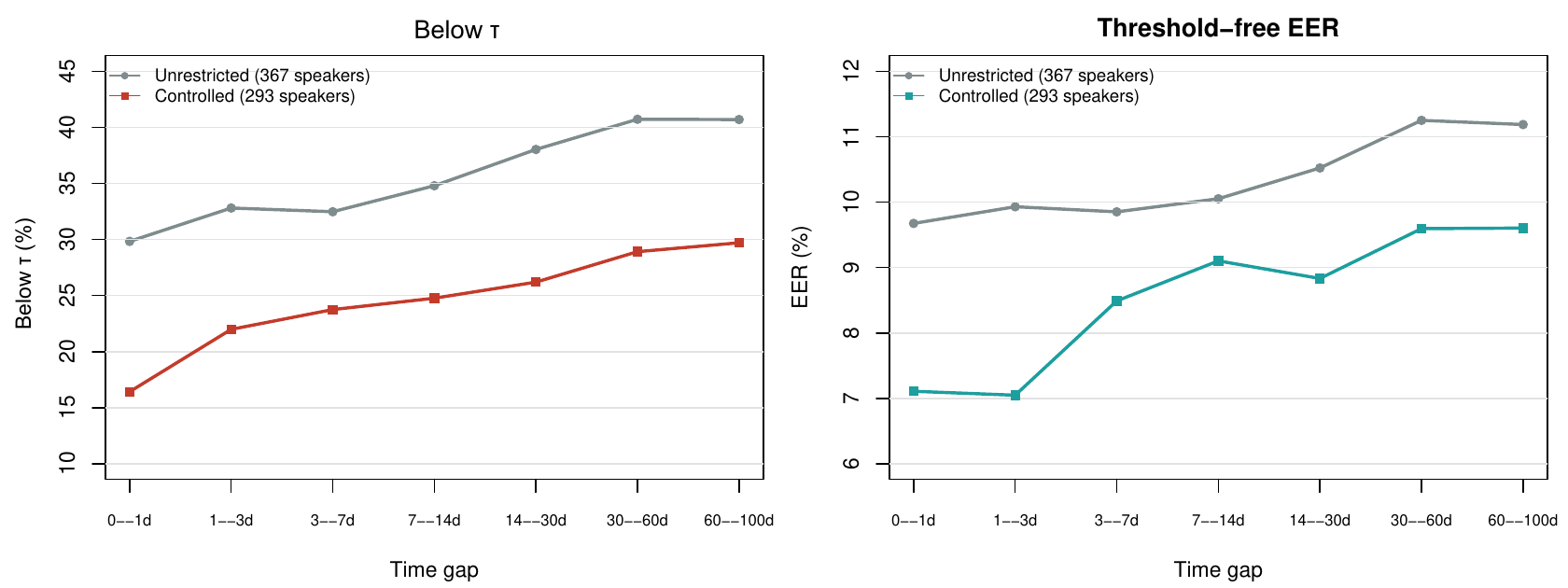}
    \caption{Below-$\tau$ and EER across time gaps in the unrestricted data and in the controlled observational subset. The controlled comparison reduces several measured context differences but does not establish a causal effect of time.}
    \label{fig:app-smiip-controlled}
\end{figure*}

\begin{figure*}[t]
    \centering
    \includegraphics[width=\textwidth]{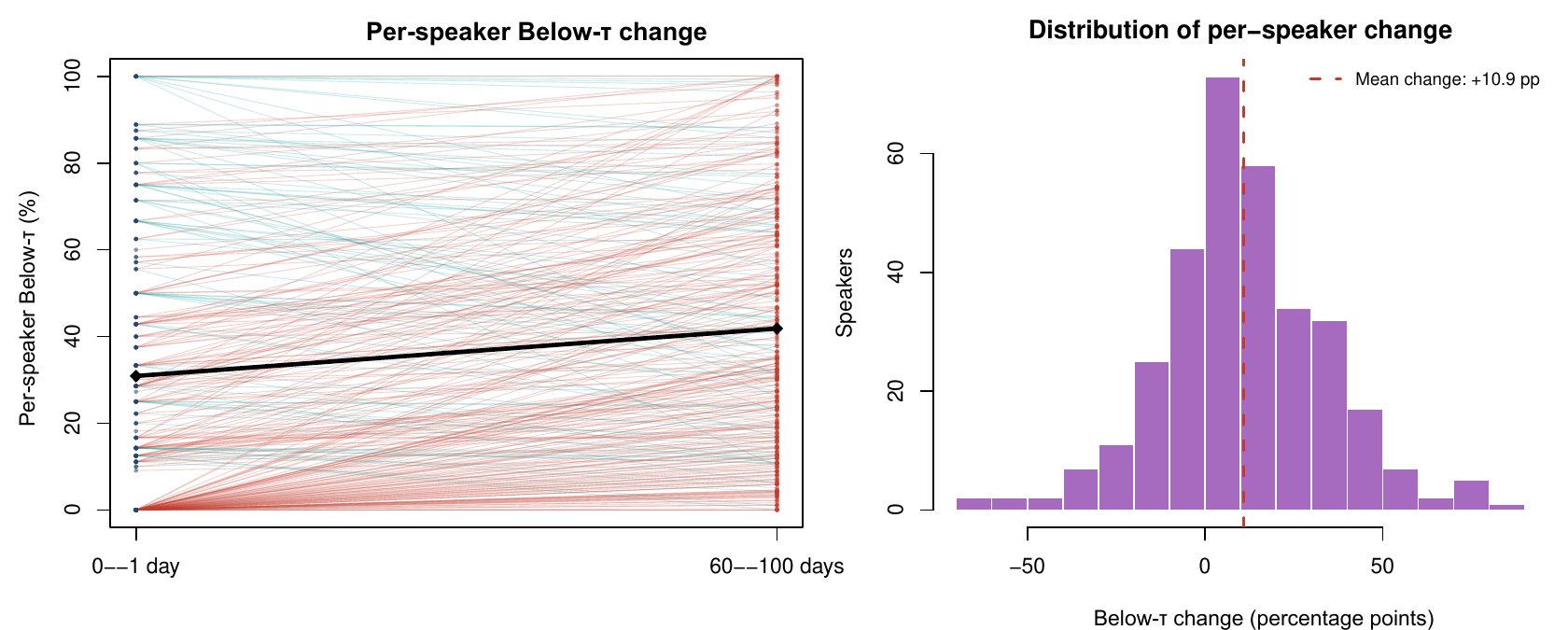}
    \caption{Speaker-level Below-$\tau$ in the nearest and farthest within-release time buckets. Lines connect observations from the same speaker where both buckets are available.}
    \label{fig:app-smiip-speaker}
\end{figure*}

\subsection{Static Verification Protocol}
\label{app:smiip-static-protocol}

SMIIP-TV contains recordings from five daily periods, with three utterances per period and metadata describing recording environments, noise, physical states, and time of day. We use text-independent utterances throughout. The approximately one-year extension includes 118 matched speakers but lacks the full context annotations of the original release.

For static verification, each speaker's earliest utterance serves as enrollment. We use only probes recorded within three days of enrollment for threshold calibration. The 98th percentile of the corresponding impostor-score distribution defines $\tau_{2\%}=0.4751$, which is frozen across subsequent time-gap analyses and causal replay policies. For bucketed EER estimation, each genuine probe is paired with 20 randomly sampled impostors. Longer-gap genuine probes and downstream labels are excluded from calibration.

Below-$\tau$ measures genuine probes rejected by the frozen enrollment template; it is distinct from the false-new user rate (FNUR), which measures erroneous entity creation during causal replay.

\subsection{Matched-Set ReDimNet2-B6 Static Robustness Check}
\label{app:redimnet2-static}

To test whether fixed-template degradation is specific to the ECAPA encoder used in the operational system, we repeat the static verification protocol with ReDimNet2-B6. The comparison is restricted
to the intersection of utterances successfully processed by both encoders, containing 205,765 text-independent SMIIP-TV utterances. ECAPA and ReDimNet2-B6 therefore use identical speakers, enrollment utterances, genuine probes, sampled impostor trials, and time buckets.

Each encoder is calibrated separately under the same protocol: the earliest available utterance is enrollment, probes within three days define the calibration window, and the threshold is selected at target FAR=2\% and then frozen for all remaining analyses. This gives $\tau_{\mathrm{ECAPA}}=0.4752$ and $\tau_{\mathrm{ReDimNet2}}=0.6131$. The threshold values themselves are encoder-specific and are not compared directly; EER and within-encoder changes in below-threshold genuine rate are the relevant cross-encoder diagnostics.

\begin{figure*}[t]
    \centering
    \begin{minipage}[t]{0.36\textwidth}
        \vspace{0pt}\centering
        \includegraphics[width=\linewidth]{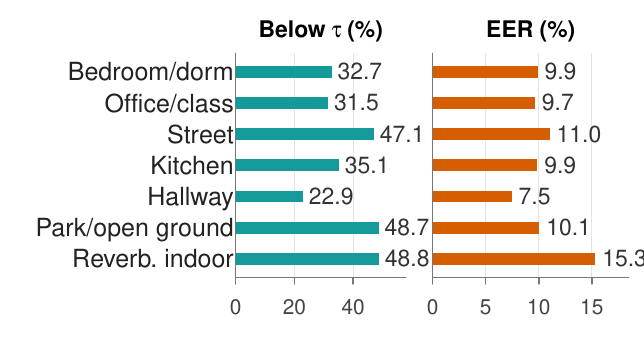}
        \par\vspace{0.2ex}\textbf{(a) Environment}
    \end{minipage}\hfill
    \begin{minipage}[t]{0.315\textwidth}
        \vspace{0pt}\centering
        \includegraphics[width=\linewidth]{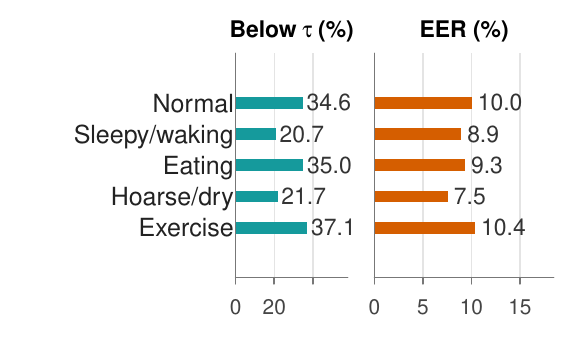}
        \par\vspace{0.2ex}\textbf{(b) Speaker state}
    \end{minipage}\hfill
    \begin{minipage}[t]{0.305\textwidth}
        \vspace{0pt}\centering
        \includegraphics[width=\linewidth]{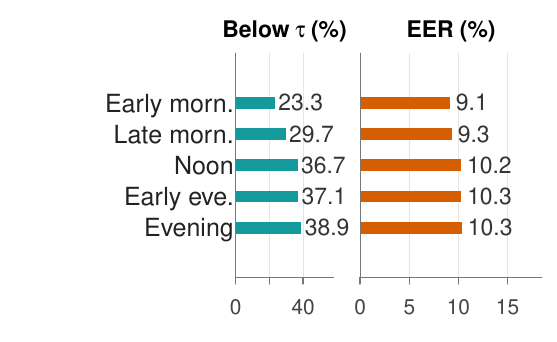}
        \par\vspace{0.2ex}\textbf{(c) Time of day}
    \end{minipage}
    \caption{Fixed-template Below-$\tau$ and EER across recording environment, speaker state, and time of day. Probes are within 14 days of enrollment, and displayed categories contain at least 300 probes. Complete category breakdowns follow below.}
    \label{fig:smiip-context}
\end{figure*}

\begin{figure*}[t]
    \centering
    \includegraphics[width=\textwidth]{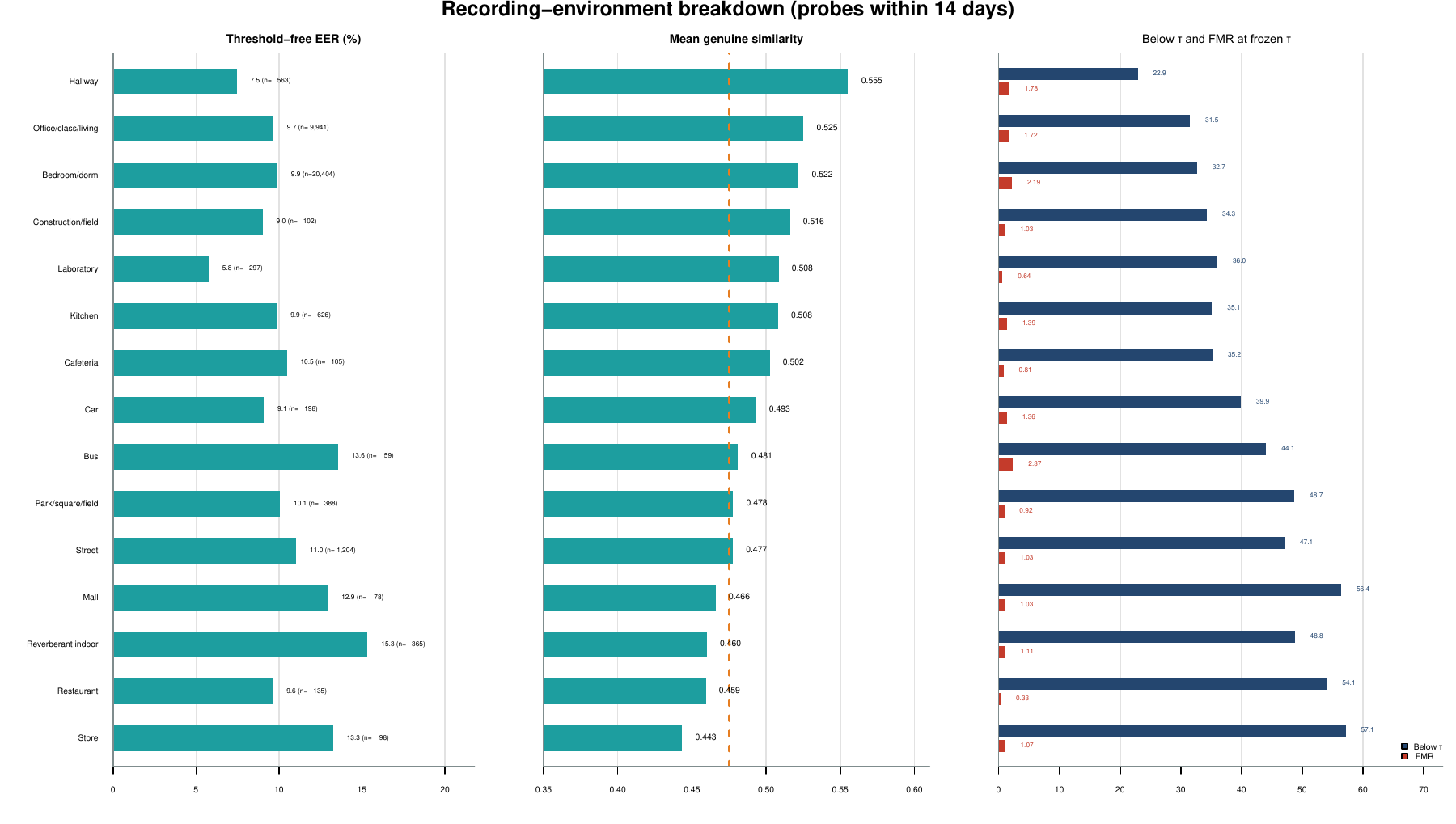}
    \caption{Complete near-enrollment breakdown by recording environment. Sample sizes are shown in the plot; categories with small support should be interpreted cautiously.}
    \label{fig:app-smiip-env}
\end{figure*}

\begin{figure*}[t]
    \centering
    \includegraphics[width=\textwidth]{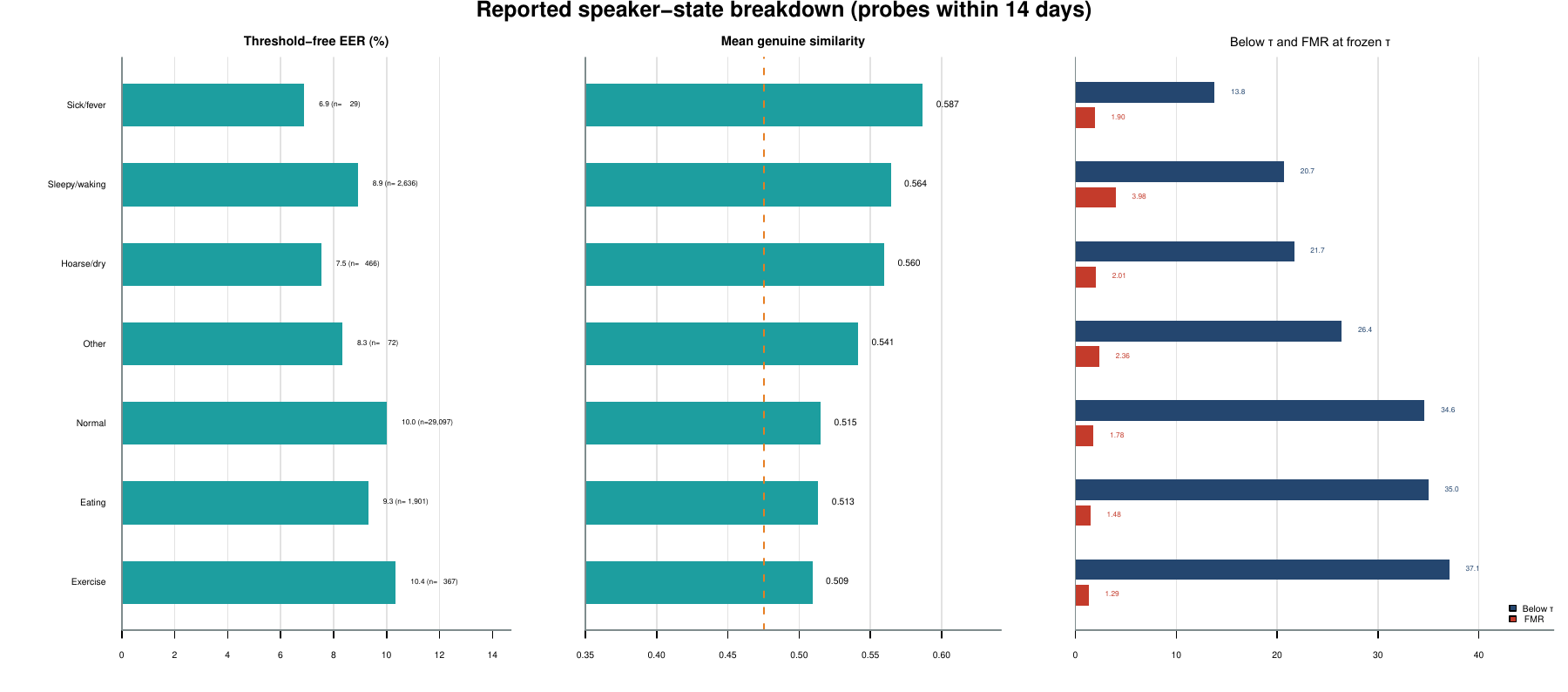}
    \caption{Complete near-enrollment breakdown by reported speaker state. Lower Below-$\tau$ in a category does not necessarily imply uniformly easier routing because impostor acceptance~(shown as FMR in the figure) and sample support can differ.}
    \label{fig:app-smiip-state}
\end{figure*}

\begin{figure*}[t]
    \centering
    \includegraphics[width=\textwidth]{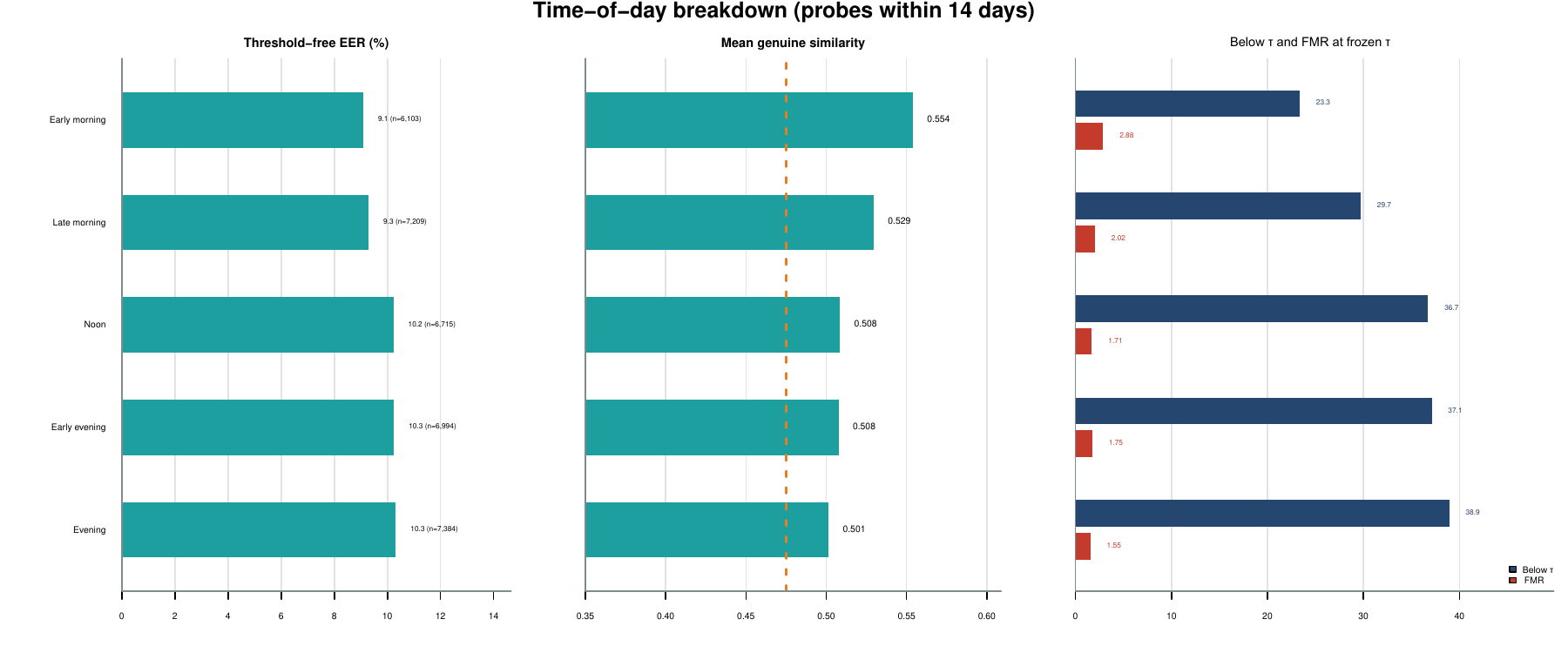}
    \caption{Complete near-enrollment breakdown by time of day. All five periods have broad speaker coverage, but the comparison remains observational.}
    \label{fig:app-smiip-tod}
\end{figure*}

\begin{table*}[t]
\centering
\small
\setlength{\tabcolsep}{4pt}
\begin{tabular}{p{0.15\textwidth}p{0.29\textwidth}p{0.31\textwidth}p{0.18\textwidth}}
\hline
Dimension & Small span (control) & Large span (higher drift) & Purpose \\
\hline
Intra-day TOD & Same day and same time-of-day bucket as enrollment & Same day, at least two time-of-day buckets from enrollment & Isolate within-day change \\
Inter-day & 0--3 days after enrollment & 30--100 days after enrollment & Short-term temporal drift \\
Long-term cross-batch & Same-release SMIIP-TV probe & Follow-up probe approximately one year later & Long-term return \\
Environment & Within 14 days, same environment as enrollment & Within 14 days, different environment & Approximate environment contrast \\
Speaker state & Within 14 days, normal state & Within 14 days, non-normal state (e.g., sleepy, eating, hoarse, or exercise) & Approximate state contrast \\
Long-term $\times$ hist.-env. & Follow-up probes from speakers with low historical environment entropy & Follow-up probes from speakers with high historical environment entropy & Long-term return by prior variability \\
\hline
\end{tabular}
\caption{Small- and large-span definitions for the six SMIIP-TV online replay dimensions. Hist.-env. uses variability in the original SMIIP-TV histories.}
\label{tab:app-smiip-spans}
\end{table*}

Table~\ref{tab:app-redimnet2-static} and Figures~\ref{fig:comp-smiip-shortterm}--\ref{fig:comp-smiip-longterm} compare the matched-set verification results. ReDimNet2-B6 consistently improves absolute verification separability. For example, its EER is 3.34\% at 0--1 days and 6.35\% at 365--450 days, compared with 9.71\% and 14.09\% for ECAPA on the same utterance set. Nevertheless, ReDimNet2-B6's below-threshold genuine rate increases from 5.19\% to 12.96\% over the same interval. Thus, stronger acoustic representations reduce the severity of longitudinal mismatch but do not make a single enrollment embedding a time-invariant entity key.

\subsection{Matched-Set ReDimNet2-B6 Causal Replay Check}
\label{app:redimnet2-replay}

We additionally test whether the online maintenance findings depend on the ECAPA representation used by the operational system. On the same common-utterance set as the static comparison, we construct causal multi-speaker replay streams once and replay the identical session JSONL streams with ECAPA-TDNN and ReDimNet2-B6 embeddings. The six drift dimensions, five random seeds, policy implementations, turn order, and causality constraint are unchanged; only the encoder and its independently calibrated frozen threshold differ.

For fragmentation, let $n_{s,k}$ count evaluated turns of gold speaker $s$ assigned to predicted entity $k$, and let $S$ be the gold speakers in a replay. We compute each speaker's fraction outside its dominant predicted identity, then macro-average across speakers:
\begin{equation}
\mathrm{Frag.}=\frac{100}{|S|}\sum_{s\in S}
\left(1-\frac{\max_k n_{s,k}}{\sum_k n_{s,k}}\right).
\end{equation}
This measures within-speaker dispersion, not cross-speaker contamination: assigning everyone to one entity can give zero fragmentation despite poor partition quality. We therefore report it alongside B$^3$ F1 and FNUR.

Table~\ref{tab:app-redimnet2-replay} isolates the encoder effect for all-exemplar KNN: mean B$^3$ F1 rises from 65.45\% with ECAPA to 73.68\% with ReDimNet2-B6, an 8.23-point improvement. ReDimNet2-B6 improves B$^3$ F1 in five of six conditions and reduces false-new decisions and fragmentation in most conditions. Together with the static verification results, this comparison shows that stronger embeddings mitigate acoustic mismatch while fixed-template degradation persists. The relative benefit of alternative maintenance policies under ReDimNet2-B6 remains unmeasured.

\subsection{Time-Gap and Score Diagnostics}
\label{app:smiip-time}

For the ECAPA operational setting used by the online replay and LoCoMo-ID experiments, Figure~\ref{fig:app-smiip-longterm} expands the three representative rows in Table~\ref{tab:smiip-static} to the full observed enrollment--probe range. The fixed threshold remains unchanged across buckets.

Figure~\ref{fig:app-smiip-controlled} compares the unrestricted data with the controlled bedroom/dormitory, normal-state, normal-noise subset described in the main text. Figure~\ref{fig:app-smiip-speaker} provides a speaker-level view of Below-$\tau$ change, showing that the aggregate trend is not attributable only to a small number of outlying users.

\subsection{Complete Context Breakdowns}
\label{app:smiip-context}

\begin{table}[t]
\centering
\resizebox{\columnwidth}{!}{\large
\begin{tabular}{lrrrrr}
\hline
Policy & 
$P_{\mathrm{create}}$ $\uparrow$ &
$R_{\mathrm{create}}$ $\uparrow$ &
$P_{\mathrm{link}}$ $\uparrow$ &
$R_{\mathrm{known}}$ $\uparrow$ &
$F_{B^3}$ $\uparrow$ \\
\hline
First exemplar & 21.77 & 52.90 & \textbf{66.25} & \textbf{62.90} & 46.82 \\
Enrollment mean & 27.23 & 42.83 & 57.18 & 56.51 & 53.58 \\
Latest exemplar & 23.95 & 54.14 & 34.60 & 33.10 & 32.20 \\
Online mean & \textbf{31.24} & 38.48 & 46.54 & 46.65 & 55.99 \\
EMA & 31.07 & 42.01 & 41.36 & 41.35 & 52.03 \\
All-exemplar KNN & 31.06 & \textbf{56.24} & 52.56 & 51.30 & \textbf{65.31} \\
\hline
\end{tabular}
}
\caption{Macro-averaged large-span decision and B$^3$ results across six SMIIP-TV conditions (five seeds; \%; higher is better).}
\label{tab:smiip-online-decisions}
\end{table}

\begin{figure*}[t]
    \centering
    \includegraphics[width=\textwidth]{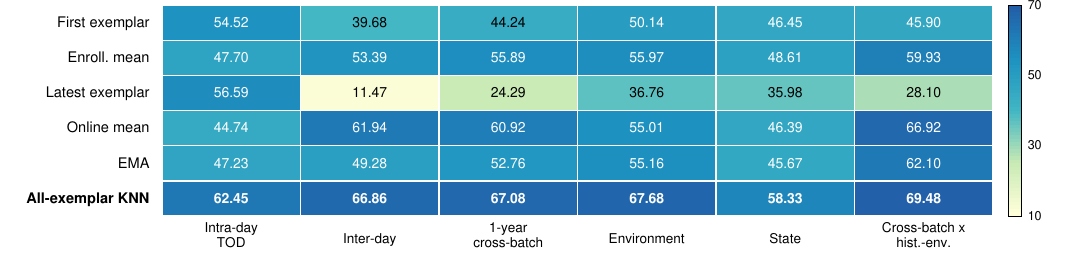}
    \vspace{-10mm}\caption{Large-span B$^3$ F1 (\%), averaged over five seeds. Bold marks the best policy per condition. TOD denotes time of day; Hist.-env. denotes historical environment variability.}\vspace{-4mm}
    \label{fig:smiip-online-bcubed}
\end{figure*}

\begin{figure*}[t]
    \centering
    \includegraphics[width=\textwidth]{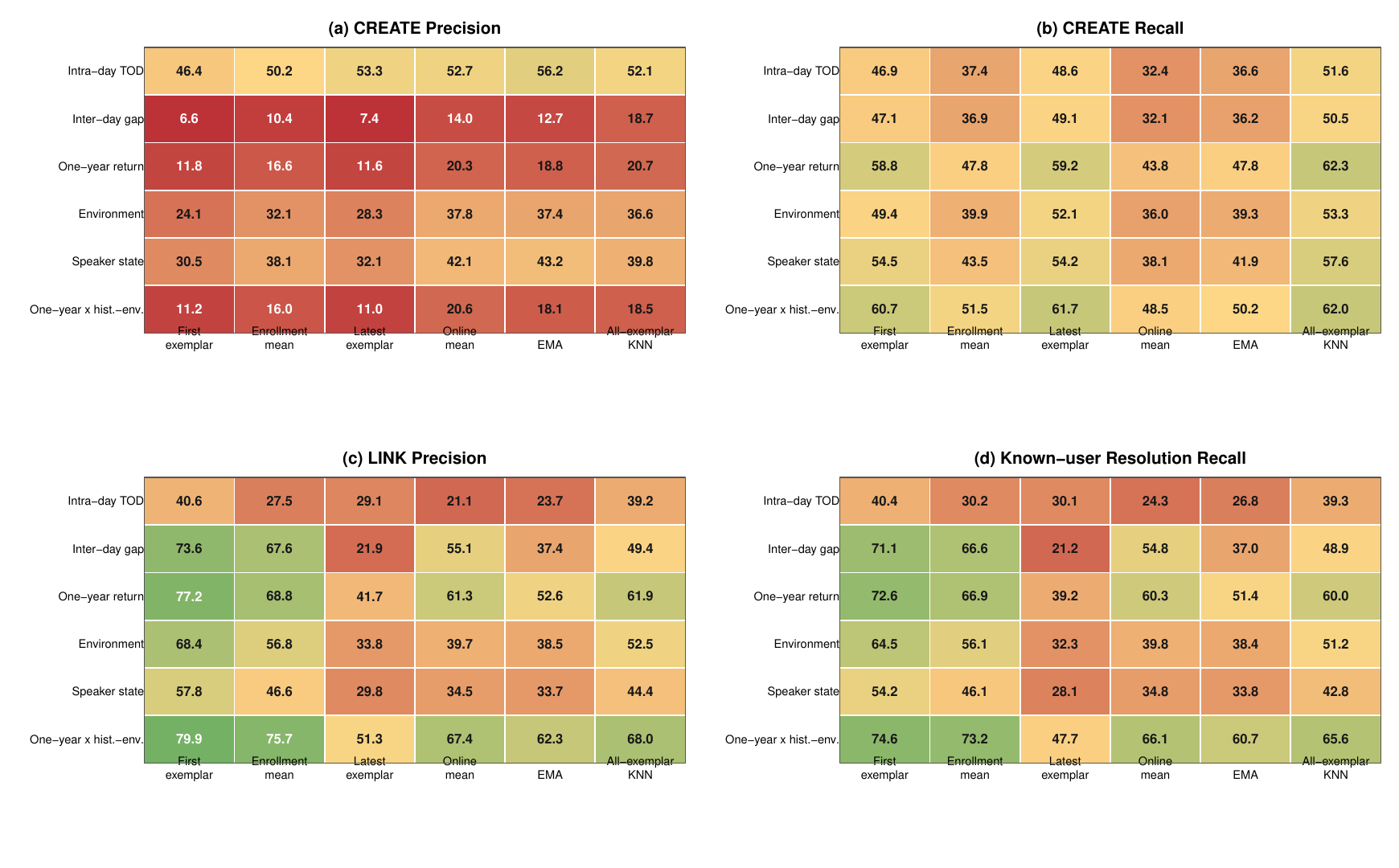}
    \caption{Large-span CREATE/LINK decision metrics across six policies and six conditions (five-seed means; \%; higher is better).}
    \label{fig:app-smiip-decisions}
\end{figure*}

Figures~\ref{fig:app-smiip-env}, \ref{fig:app-smiip-state}, and \ref{fig:app-smiip-tod} report all available environment, physical-state, and time-of-day categories, including the lower-count categories omitted from Figure~\ref{fig:smiip-context}. Each figure jointly displays threshold-free EER, mean genuine similarity, and fixed-threshold Below-$\tau$ and impostor acceptance rate. Several categories contain few speakers or probes; these plots are descriptive diagnostics and should not be interpreted as controlled causal comparisons.

\subsection{Online Replay Conditions}
\label{app:smiip-spans}

Table~\ref{tab:app-smiip-spans} defines the low-drift control and higher-drift treatment used for each causal replay. ``Small'' and ``large'' refer to the condition span relative to enrollment, not the number of speakers, turns, or sessions. For each speaker and span, the replay begins with enrollment observations and then consumes probes from the corresponding span in chronological order. Small- and large-span replays are constructed independently rather than concatenated into a single stream. Multiple speaker queues are randomly interleaved while preserving per-speaker first-in, first-out~(FIFO) order. We do not impose session-level domain consistency: different speakers in the same synthetic session may occupy different stages of their own trajectories.

For each replay, we randomly interleave speaker queues while preserving their chronological order, selecting one to three consecutive turns from a speaker at each step. We evaluate six training-free representation policies: First exemplar, Enrollment mean, Latest exemplar, Online mean, exponential moving average (EMA; $\alpha=0.3$), and All-exemplar KNN. The first five retain or update a single speaker template, whereas All-exemplar KNN stores accepted historical embeddings and averages the top-five cosine similarities for identity matching. All policies use the same frozen encoder and assignment threshold. Metrics are averaged over five seeds within each condition and then macro-averaged across the six conditions.

\subsection{Online Policy Diagnostics}
\label{app:smiip-online}

Table~\ref{tab:smiip-online-decisions} summarizes the macro-averaged causal replay results across six conditions. Figure~\ref{fig:smiip-online-bcubed} provides the corresponding per-condition B$^3$ F1 comparison, while Figure~\ref{fig:app-smiip-decisions} reports detailed \textsc{create}/\textsc{link} metrics. These diagnostics illustrate the trade-off between speaker fragmentation and incorrect identity linking. Low FNUR alone is insufficient: a policy may link returning turns frequently while still selecting the wrong target. The decision matrices should therefore be interpreted jointly with the B$^3$ results in Figure~\ref{fig:smiip-online-bcubed}.

\section{Additional LoCoMo-ID Routing and Private-Access Analyses}
\label{app:locomo-private}

\begin{table*}[t]
\centering
\small
\resizebox{\columnwidth}{!}{%
\begin{tabular}{lrr}
\hline
Acoustic evidence / policy & Clean & Condition-shifted \\
\hline
ECAPA / Fixed & 65.53 & 61.09 \\
ECAPA / Update & 66.63 & 65.54 \\
ECAPA / Gated update & 66.87 & 64.99 \\
ECAPA / All-exemplar KNN & 66.65 & 66.25 \\
Parakeet+ECAPA / Fixed & 66.30 & 65.36 \\
Parakeet+ECAPA / Update & 65.93 & 65.74 \\
Parakeet+ECAPA / Gated update & 66.03 & 66.00 \\
Parakeet+ECAPA / All-exemplar KNN & 65.88 & 65.68 \\
\hline
\end{tabular}}
\caption{Full-set all-memory QA (\%) for eight policies on clean and condition-shifted speech. Each condition comprises 30 collections and 5,958 query audios. Scores are micro-pooled judge-correct rates without speaker-based retrieval filtering.}
\label{tab:app-locomo-all-memory-full}
\end{table*}

\begin{table*}[t]
\centering
\small
\resizebox{\textwidth}{!}{\begin{tabular}{llrrrrrrr}
\hline
Condition & Acoustic evidence / policy & Turns & Entities & $P_{\mathrm{create}}$ & $R_{\mathrm{create}}$ & $P_{\mathrm{link}}$ & $R_{\mathrm{known}}$ & FNUR \\
\hline
Clean & ECAPA / Fixed & 17643 & 193 & 30.05 & 96.67 & 99.98 & 99.22 & 0.77 \\
 & ECAPA / Update & 17643 & 106 & 54.72 & 96.67 & 97.49 & 97.23 & 0.27 \\
 & ECAPA / Gated update & 17643 & 113 & 51.33 & 96.67 & 99.95 & 99.65 & 0.31 \\
 & ECAPA / All-exemplar KNN & 17643 & 95 & 61.05 & 96.67 & 97.48 & 97.29 & 0.21 \\
 & Parakeet+ECAPA / Fixed & 17644 & 69 & 79.71 & 91.67 & 81.41 & 81.36 & 0.08 \\
 & Parakeet+ECAPA / Update & 17644 & 67 & 82.09 & 91.67 & 81.40 & 81.37 & 0.07 \\
 & Parakeet+ECAPA / Gated update & 17644 & 67 & 82.09 & 91.67 & 81.40 & 81.37 & 0.07 \\
 & Parakeet+ECAPA / All-exemplar KNN & 17644 & 67 & 82.09 & 91.67 & 81.37 & 81.34 & 0.07 \\
\hline
Condition-shifted & ECAPA / Fixed & 17644 & 410 & 14.15 & 96.67 & 99.63 & 97.65 & 2.00 \\
 & ECAPA / Update & 17644 & 216 & 26.85 & 96.67 & 97.30 & 96.43 & 0.90 \\
 & ECAPA / Gated update & 17644 & 224 & 25.89 & 96.67 & 99.67 & 98.74 & 0.94 \\
 & ECAPA / All-exemplar KNN & 17644 & 154 & 37.66 & 96.67 & 97.48 & 96.96 & 0.55 \\
 & Parakeet+ECAPA / Fixed & 17643 & 79 & 67.09 & 88.33 & 83.26 & 83.17 & 0.15 \\
 & Parakeet+ECAPA / Update & 17643 & 70 & 75.71 & 88.33 & 83.26 & 83.22 & 0.10 \\
 & Parakeet+ECAPA / Gated update & 17643 & 70 & 75.71 & 88.33 & 83.25 & 83.20 & 0.10 \\
 & Parakeet+ECAPA / All-exemplar KNN & 17643 & 69 & 76.81 & 88.33 & 83.11 & 83.07 & 0.09 \\
\hline
\end{tabular}}
\caption{History turn-micro causal routing results for all eight policies under clean speech and condition shifted at merge threshold 0.489. Values are percentages except Turns and Entities.}
\label{tab:app-locomo-routing}
\end{table*}

\begin{table*}[t]
\centering
\small
\resizebox{\textwidth}{!}{\begin{tabular}{llrrrrrr}
\hline
Condition & Acoustic evidence / policy & C & W & D & Safe auth. & QA & QA$\mid$Correct \\
\hline
Clean & ECAPA / Fixed & 4084 & 0 & 29 & 99.29 & 71.41 & 71.91 \\
 & ECAPA / Update & 3993 & 115 & 5 & 97.08 & 73.43 & 75.63 \\
 & ECAPA / Gated update & 4110 & 0 & 3 & 99.93 & 75.27 & 75.33 \\
 & ECAPA / All-exemplar KNN & 3993 & 116 & 4 & 97.08 & 73.64 & 75.86 \\
 & Parakeet+ECAPA / Fixed & 3833 & 273 & 7 & 93.19 & 68.08 & 73.05 \\
 & Parakeet+ECAPA / Update & 3833 & 272 & 8 & 93.19 & 68.25 & 73.23 \\
 & Parakeet+ECAPA / Gated update & 3833 & 272 & 8 & 93.19 & 68.25 & 73.23 \\
 & Parakeet+ECAPA / All-exemplar KNN & 3833 & 274 & 6 & 93.19 & 68.10 & 73.08 \\
\hline
Condition-shifted & ECAPA / Fixed & 4079 & 1 & 33 & 99.17 & 60.15 & 60.65 \\
 & ECAPA / Update & 3993 & 115 & 5 & 97.08 & 70.24 & 72.35 \\
 & ECAPA / Gated update & 4109 & 0 & 4 & 99.90 & 71.77 & 71.84 \\
 & ECAPA / All-exemplar KNN & 3993 & 116 & 4 & 97.08 & 72.40 & 74.58 \\
 & Parakeet+ECAPA / Fixed & 3560 & 318 & 235 & 86.55 & 63.36 & 73.20 \\
 & Parakeet+ECAPA / Update & 3786 & 320 & 7 & 92.05 & 67.76 & 73.61 \\
 & Parakeet+ECAPA / Gated update & 3739 & 320 & 54 & 90.91 & 67.20 & 73.92 \\
 & Parakeet+ECAPA / All-exemplar KNN & 3787 & 321 & 5 & 92.07 & 67.57 & 73.38 \\
\hline
\end{tabular}}
\caption{Speaker-sensitive Top-1 authorization and end-to-end answer utility over 4,113 questions per condition (clean and condition-shifted). C/W/D are counts, other values are percentages.}
\label{tab:app-locomo-access}
\end{table*}

\begin{table*}[t]
\centering
\small
\setlength{\tabcolsep}{3pt}
\resizebox{\textwidth}{!}{%
\begin{tabular}{lrrrrrrr}
\hline
\multicolumn{1}{c}{Policy}
& \multicolumn{1}{c}{\shortstack{Online\\ID}}
& \multicolumn{1}{c}{$F_{B^3}$}
& \multicolumn{1}{c}{C / W / D}
& \multicolumn{1}{c}{\shortstack{Targeted\\strict (19)}}
& \multicolumn{1}{c}{\shortstack{Targeted\\raw (20)}}
& \multicolumn{1}{c}{\shortstack{Full strict\\(1,371)}}
& \multicolumn{1}{c}{\shortstack{QA$\mid$\\Correct}} \\
\hline
\multicolumn{8}{l}{\textit{ECAPA}} \\
Fixed & 98.84 & 96.54 & 1342/18/11 & 15/19 (78.95) & 15/20 (75.00) & 930/1371 (67.83) & 69.30 \\
Update & 93.35 & 95.45 & 1335/32/4 & 15/19 (78.95) & 16/20 (80.00) & 988/1371 (72.06) & 74.01 \\
Gated update & 96.25 & 97.77 & 1368/0/3 & 15/19 (78.95) & 15/20 (75.00) & 1009/1371 (73.60) & 73.76 \\
KNN & 94.85 & 97.20 & 1335/33/3 & 15/19 (78.95) & 16/20 (80.00) & 987/1371 (71.99) & 73.93 \\
\hline
\multicolumn{8}{l}{\textit{Parakeet+ECAPA}} \\
Fixed & 74.03 & 72.86 & 1335/32/4 & 13/19 (68.42) & 14/20 (70.00) & 980/1371 (71.48) & 73.41 \\
Update & 71.96 & 72.57 & 1333/32/6 & 14/19 (73.68) & 15/20 (75.00) & 991/1371 (72.28) & 74.34 \\
Gated update & 73.87 & 73.10 & 1334/32/5 & 13/19 (68.42) & 14/20 (70.00) & 990/1371 (72.21) & 74.21 \\
KNN & 71.93 & 72.60 & 1335/32/4 & 14/19 (73.68) & 15/20 (75.00) & 992/1371 (72.36) & 74.31 \\
\hline
\end{tabular}}
\caption{Complete four-speaker results on ten clean collections. Targeted strict QA (19) and full strict QA (1,371) require correct authorization and a judge-correct answer; targeted raw judge (20) requires only a judge-correct answer and includes the non-gated item. C/W/D are full-set authorization counts. QA$\mid$Correct conditions full strict QA on C. All rates are percentages; KNN denotes All-exemplar KNN.}
\label{tab:app-locomo-interleaved}
\end{table*}

\begin{table*}[t]
\centering
\small
\begin{tabular}{rrrrrrr}
\hline
$\tau_{\mathrm{update}}$ & Entities & Online (\%) & B$^3$ F1 (\%) & Query assigned / 5958 & W / 4113 & QA (\%) \\
\hline
0.50 & 106 & 97.23 & 98.41 & 5953 & 115 & 73.33 \\
0.54 & 110 & 98.65 & 99.11 & 5954 & 32 & 74.81 \\
0.58 & 113 & 99.64 & 99.54 & 5955 & 0 & 75.27 \\
0.62 & 114 & 99.64 & 99.54 & 5955 & 0 & 75.27 \\
0.66 & 114 & 99.64 & 99.55 & 5955 & 0 & 75.47 \\
0.70 & 118 & 99.61 & 99.45 & 5955 & 0 & 75.44 \\
\hline
\end{tabular}
\caption{Clean ECAPA gated-update sweep, fixed merge threshold 0.489. Routing/cluster scores and query assigned are from the upstream replay; W and QA are from the separate complete downstream threshold-sweep runs, not the default-run main table.}
\label{tab:app-locomo-threshold}
\end{table*}

\begin{table*}[t]
\centering
\small
\begin{tabular}{llrrr}
\hline
Condition & ECAPA policy & $<1$ s: correct / $n$ & $1$--$<2$ s: correct / $n$ & $\geq 2$ s: correct / $n$ \\
\hline
Clean & Gated update & 48 / 59 (81.36\%) & 375 / 382 (98.17\%) & 17157 / 17202 (99.74\%) \\
 & All-exemplar KNN & 50 / 59 (84.75\%) & 372 / 382 (97.38\%) & 16742 / 17202 (97.33\%) \\
Condition-shifted & Gated update & 2 / 13 (15.38\%) & 195 / 252 (77.38\%) & 17223 / 17379 (99.10\%) \\
 & All-exemplar KNN & 3 / 13 (23.08\%) & 210 / 252 (83.33\%) & 16894 / 17379 (97.21\%) \\
\hline
\end{tabular}
\caption{Post-hoc duration-bucket first-seen-binding accuracy on evaluated history turns. Each policy/condition denominator sums to 17,643 (clean) or 17,644 (condition-shifted); the short bins are very small in the augmented condition.}
\label{tab:app-locomo-duration}
\end{table*}

\begin{table*}[t]
\centering
\small
\setlength{\tabcolsep}{5pt}
\begin{tabular}{p{0.25\textwidth}p{0.30\textwidth}p{0.35\textwidth}}
\hline
Stage & Model & Decoding configuration \\
\hline
History and query transcription & Qwen3-Omni-30B-A3B-Instruct & Sampling: temperature 0.7, top-$p$ 0.8, top-$k$ 20; maximum 512 new tokens. \\
LightMem factual extraction & Qwen3-Omni-30B-A3B-Instruct & Sampling: temperature 0.1, top-$p$ 0.1; maximum 16,000 new tokens. The configured top-$k=1$ is not forwarded to the decoder. \\
Answer generation & Qwen3-Omni-30B-A3B-Instruct & Greedy decoding (\texttt{do\_sample=False}); maximum 300 new tokens. \\
LLM-as-a-judge & Qwen3-Omni-30B-A3B-Instruct & Greedy decoding (\texttt{do\_sample=False}); maximum 200 new tokens. \\
Text embedding & bge-small-en-v1.5 & 384-dimensional embedding; no generative decoding. \\
\hline
\end{tabular}
\caption{Models and generation settings for the reported LoCoMo-ID experiments.}
\label{tab:implementation-config}
\end{table*}

\subsection{Full-Set All-Memory QA}
\label{app:locomo-all-memory-full}

Table~\ref{tab:app-locomo-all-memory-full} reports unrestricted-retrieval answer accuracy on all 5,958 query audios per condition for the eight policies in Table~\ref{tab:locomo-all-qa}. The scores combine 4,113 sensitive-query results from all-memory evaluation with 1,845 non-sensitive-query results, which also used unrestricted retrieval in the corresponding full-set runs.

\subsection{Full Causal Decision and Private-Access Results}

Table~\ref{tab:app-locomo-routing} reports additional history-routing metrics and entity counts. Entities is the total number of predicted entities across 30 independent collection-local registries; all rate metrics are percentages.

Table~\ref{tab:app-locomo-access} reports the complete Top-1 access outcomes. ``Safe auth.'' is the correctly authorized fraction, $100C/N_q$, with $N_q=4{,}113$. Wrong authorization is counted even if the LLM subsequently refuses or answers incorrectly. QA divides the correct-authorized and judge-correct count by 4,113; conditional QA divides by $C$. A query rejected before retrieval (D) differs from a downstream refusal on an authorized query.

Identity filtering need not improve answer accuracy alone. Fixed's augmented all-memory QA is 69.27\%, versus 60.15\% strict identity-gated QA. Its fragmented histories (410 entities for 60 gold speakers) are consistent with the selected Top-1 partition retaining only part of the requester's evidence. This is a possible mechanism, not an isolated causal test: the two QA metrics also impose different authorization requirements. Parakeet+ECAPA shows the complementary failure mode. Despite fewer entities, its lower LINK precision and poorer B$^3$ F1 are consistent with incorrect associations across speakers, so entity count alone cannot establish correct access. These observations motivate joint reporting of partition structure, authorization, and answer utility.

\subsection{Four-Speaker Interleaved Query Results}
\label{app:locomo-four-speaker}

Table~\ref{tab:app-locomo-interleaved} expands the four-speaker results in the main text. Each policy uses its completed four-speaker history and memory state. Full identity-gated QA divides the number of correctly authorized, judge-correct answers by 1,371, while conditional QA divides that number by $C$. Targeted strict QA uses the 19 identity-gated questions; targeted raw-judge QA evaluates answer correctness on all 20 questions without requiring correct authorization, including the remaining verbatim question.


\subsection{Update-Threshold and Utterance-Duration Diagnostics}
\label{app:locomo-sensitivity}

We vary the center-update threshold on clean ECAPA speech while holding the merge and query lookup threshold at 0.489 (Table~\ref{tab:app-locomo-threshold}). We select 0.58 as the lowest tested threshold in the 0.58--0.66 range where online accuracy and B$^3$ F1 are nearly stable and no wrong authorizations occur; lower values (0.50 and 0.54) produce 115 and 32 wrong authorizations, respectively.

For a post-hoc utterance-length analysis, we join the recorded \texttt{duration\_seconds} and chronological per-turn assignments for the two global-ECAPA policies with private gold labels \emph{only after replay}. Within each of the 30 collections, a newly created local ID is immutably bound to the gold identity of its first evaluated turn, and a first-seen gold speaker scores as correct only if a new local ID is created (Section~\ref{sec:online-metrics}). Each row of Table~\ref{tab:app-locomo-duration} counts actual waveform duration, not lexical length, and reports micro-averaged first-seen-binding accuracy; rates are not a controlled intervention on duration. Shifted condition changes waveform durations and thus the bucket populations relative to clean speech.

All 60 first-appearance gold turns in each condition fall in the $\geq 2$ s bin. In particular, 11 of 13 condition-shifted $<1$ s gated turns are erroneous duplicate \textsc{creates}; these estimates are high variance and cannot establish a general short-speech failure rate.


\section{Backend Models and Decoding Configuration}
\label{app:backend-config}

All LoCoMo-ID experiments use local \texttt{Qwen3-Omni-30B-A3B-Instruct} for history/query transcription, LightMem factual extraction, answer generation, and LLM-as-a-judge evaluation. The speech-generation branch is disabled. LightMem uses \texttt{bge-small-en-v1.5} (384 dimensions) for both indexing memory entries and embedding text queries; Qdrant provides the vector store. Before fact extraction, we use the multilingual MeetingBank checkpoint from LLMLingua-2 for pre-compression (rate 0.6) and topic segmentation. The factual-extraction, answer, and judge prompts are otherwise unchanged from LightMem and LoCoMo.

Table~\ref{tab:implementation-config} lists the model and decoding settings for these components.

\paragraph{Timing scope.}
The reported 65.6\,ms per turn measures ECAPA embedding extraction only. It excludes routing, transcription, and textual-memory construction, and does not establish full-pipeline real-time performance.

\renewenvironment{quote}
  {\list{}{\rightmargin\leftmargin\topsep=3pt\partopsep=0pt}\item\relax}
  {\endlist}
\section{LoCoMo Acoustic Tool-Use Prompt}
\label{app:locomo-prompts}

We retain LightMem's original factual-extraction prompt, the original LoCoMo answer prompt, and the original LoCoMo judge prompt unchanged, and therefore do not reproduce them here. The spoken adaptation adds the audio-transcription prompt below and a deterministic acoustic tool-use prompt for each chronologically arriving history segment. For routing, the model's role is only to invoke the tool; it does not select a speaker identity. The \texttt{extract\_segment\_embedding} tool extracts an ECAPA-TDNN embedding, scores it against the persistent registry, applies the merge threshold, and creates or links a local speaker entity. For fixed, update, and gated-update policies, only the deterministic center-maintenance rule differs; the all-exemplar policy instead maintains a history of accepted embeddings and uses mean top-five similarity. Session-log-assisted policies first group turns with local Parakeet speaker labels and then use ECAPA acoustic evidence for cross-session association. No generated transcript is used to decide identity.

\subsection{Audio Transcription Prompt}

For every history and query clip, Qwen3-Omni receives the audio together with the following system prompt and fixed user message. The resulting transcript is passed to the unchanged LightMem/LoCoMo components.

\noindent\textbf{System prompt}
\begin{quote}
\small\ttfamily
You are a careful speech transcription assistant. You will listen to one audio clip containing a question someone is asking. Transcribe EXACTLY what they say, as faithfully and completely as possible (keep every concrete name, place, date, number mentioned). Output ONLY the transcript text, nothing else — no preamble, no quotation marks, no commentary.
\end{quote}

\noindent\textbf{User message}
\begin{quote}
\small\ttfamily
[audio clip]
Transcribe this audio clip now.
\end{quote}

\subsection{Deterministic Acoustic Tool-Use Prompt}

\begin{quote}
\small\ttfamily
You are a careful speech analysis assistant building a persistent speaker memory for a conversation one audio segment at a time.

For the CURRENT audio segment, you MUST call ``extract\_segment\_embedding'' exactly once with the supplied audio path, start\_time, and end\_time. This acoustic tool extracts the segment's voiceprint, compares it with the persistent speaker registry, and automatically determines whether to link the segment to an existing local speaker or create a new one according to its configured similarity threshold and update policy.

Do NOT infer, choose, name, override, or guess the speaker identity yourself. Do NOT call any other speaker-routing tool. After the tool returns its result, reply only with a short confirmation that the current segment has been processed. Do not output a multiple-choice answer or any additional analysis.
\end{quote}

For a segment with duration $d$, the accompanying user message is:
\begin{quote}
\small\ttfamily
Audio file: <audio\_path>

Current segment: start\_time=0, end\_time=<d>.

Process this segment with ``extract\_segment\_embedding'' now.
\end{quote}

\section{LoCoMo-ID Construction Details}
\label{app:synthetic-prompts}

LoCoMo histories and reference answers are retained without LLM generation; turn selection and query rewriting use deterministic rules. The construction pipeline's only natural-language model instruction is the Qwen3-TTS VoiceDesign prompt for the anonymous voice bank.

\paragraph{Synthesis QC and augmentation.}
Each clip is initially synthesized and checked for UTMOS $\geq 3.0$ and reference-voice cosine similarity $\geq 0.489$. A failing clip receives at most four additional attempts. If none passes both checks, the highest-UTMOS candidate among the initial synthesis and retries is retained; no dialogue turn is discarded by this fallback. The procedure targets perceptual quality and intended speaker attribution, not downstream QA. Generated-to-reference similarity is not the online score against a first-turn or evolving entity center, so passing QC does not guarantee a correct routing decision. Nevertheless, using ECAPA for both QC and routing can affect clean-condition difficulty. Condition shifts are applied after clean construction without further UTMOS or reference-similarity filtering; its difficult acoustic observations are therefore not removed by this QC stage. Pre-regeneration score records were not retained, so the current experiments do not establish the before/after score distribution.

\subsection{VoiceDesign Instruction and Enrollment}

The VoiceDesign instruction follows the template:
\begin{quote}
\small\ttfamily
A \{pitch\}-pitched voice speaking at a \{speaking\_rate\} pace, with a \{timbre\} timbre and a \{expression\} expression, \{accent\} accent. \{delivery\_style\}
\end{quote}
Candidate profiles combine pitch, speaking rate, timbre, expression, English accent, and delivery style. An instantiated instruction is:
\begin{quote}
\small\ttfamily
A medium-high-pitched voice speaking at a slightly slow pace, with a bright and clear timbre and an animated and expressive expression, Scottish English accent. A composed, understated style with minimal pitch variation.
\end{quote}
ECAPA-TDNN diversity filtering retains 50 of 120 candidate voices; descriptive attributes do not necessarily yield distinct acoustics.

Each candidate voice reads the same canonical enrollment sentence:
\begin{quote}
\small\ttfamily
Hi there, this is my voice. I'm looking forward to our conversation today.
\end{quote}
Identical enrollment content controls lexical variation in ECAPA-TDNN comparisons.

\subsection{Voice-Cloning Inputs}

VoiceDesign generates each identity's enrollment audio. Qwen3-TTS Base then synthesizes history turns and queries in voice-cloning mode with these inputs:
\begin{quote}
\small\ttfamily
Reference audio: [voice-id]\_enrollment.wav\\
Reference transcript: Hi there, this is my voice. I'm looking forward to our conversation today.\\
Target text: [one LoCoMo history turn or one rewritten query]
\end{quote}
Reusing enrollment audio and text is intended to preserve identity across sessions.

\subsection{Deterministic Query Transformations}

Question rewriting is rule based rather than prompt driven. For \texttt{identity\_gated} questions, a named participant is converted to the first person, for example, ``What did Caroline research?'' becomes ``What did I research?'', and ``What is Caroline's identity?'' becomes ``What is my identity?'' For symmetric \texttt{we\_shared} questions, the two participants are converted to the first-person plural: ``What do Jon and Gina both have in common?'' becomes ``What do we have in common?'', while ``What did Jon and Gina compare their entrepreneurial journeys to?'' becomes ``What did we compare our entrepreneurial journeys to?'' Questions that cannot be transformed safely, together with adversarial questions, remain \texttt{verbatim}; for example, ``What nickname does Nate use for Joanna?'' is unchanged. Unsafe transformations fall back to their original wording.

\subsection{Session-Level Recording-Condition Augmentation}
\label{app:recording-shifts}

For each conversation instance and session, we deterministically assign one acoustic profile that specifies the room geometry, source--microphone distance and azimuth, reverberation time, microphone directivity, device response, automatic gain control, and optional environmental noise at a fixed signal-to-noise ratio. All utterances within the same session share this profile, whereas profiles can change across sessions, reflecting realistic transitions across different recording scenarios.

Each waveform is processed by room impulse response convolution, device-bandwidth filtering, gain control, noise mixing, and peak normalization. This augmentation preserves the original text, speaker--voice binding, turn order, and query audio, thereby isolating robustness to heterogeneous historical recording conditions from semantic or identity changes.

\end{document}